\documentclass[
 reprint,
 superscriptaddress,
 preprintnumbers,
 amsmath
 amssymb,
 prx,
 floatfix,
]{revtex4-2}

\DeclareMathAlphabet{\mathsfit}{T1}{\sfdefault}{\mddefault}{\sldefault}
\SetMathAlphabet{\mathsfit}{bold}{T1}{\sfdefault}{\bfdefault}{\sldefault}

\usepackage{amsmath}
\usepackage{graphicx}
\graphicspath{{fig/}}   
\usepackage{dcolumn}
\usepackage{bm}
\usepackage{makecell}
\usepackage{braket} 
\usepackage{siunitx}
\usepackage{color}
\usepackage{array}
\usepackage{xr}
\usepackage{upgreek}
\usepackage[colorlinks]{hyperref}
\usepackage[capitalize]{cleveref}
\usepackage[caption=false]{subfig}
\usepackage{multirow} 
\usepackage[T1]{fontenc}

\begin{document} 

\title{Remote Entangling Gates for Spin Qubits in Quantum Dots \\ using a Charge-Sensitive Superconducting Coupler}

\def\RLEaffil{Research Laboratory of Electronics, Massachusetts Institute of Technology, Cambridge, MA 02139, USA}
\def\LLaffil{Lincoln Laboratory, Massachusetts Institute of Technology, Lexington, MA 02421, USA}
\def\Physaffil{Department of Physics, Massachusetts Institute of Technology, Cambridge, MA 02139, USA}
\def\EECSaffil{Department of Electrical Engineering and Computer Science, Massachusetts Institute of Technology, Cambridge, MA 02139, USA}

\author{Harry Hanlim Kang}
\email{khanlim@mit.edu}
\affiliation{\RLEaffil}
\affiliation{\EECSaffil}

\author{Ilan~T.~Rosen}
\affiliation{\RLEaffil}

\author{Max~Hays}
\affiliation{\RLEaffil}

\author{Jeffrey~A.~Grover}
\affiliation{\RLEaffil}

\author{William~D.~Oliver}
\email{william.oliver@mit.edu}
\affiliation{\RLEaffil}
\affiliation{\EECSaffil}
\affiliation{\Physaffil}

\date{\today}

\begin{abstract}
We propose a method to realize microwave-activated CZ gates between two remote spin qubits in quantum dots using a charge-sensitive superconducting coupler. The qubits are longitudinally coupled to the coupler, so that the transition frequency of the coupler depends on the logical qubit states; a capacitive network model using first-quantized charge operators is developed to illustrate this. Driving the coupler transition then implements a conditional phase shift on the qubits.
Two pulsing schemes are investigated: a rapid, off-resonant pulse with constant amplitude, and a pulse with envelope engineering that incorporates dynamical decoupling to mitigate charge noise. 
We develop non-Markovian time-domain simulations to accurately model gate performance in the presence of $1/f^\beta$ charge noise.
Simulation results indicate that a CZ gate fidelity exceeding 90\% is possible with realistic parameters and noise models.
\end{abstract}

\maketitle

\section{\label{sec:1} Introduction}

Spin qubits based on electrons confined in gate-defined quantum dots constitute a promising platform for building a quantum computer~\cite{Burkard23}.
A challenge for the scalability of quantum dot architectures is the density of electrostatic gate electrodes needed to control adjacent quantum dots, which are typically fabricated at sub-100~nm pitch.
A modular architecture, where clusters of quantum dots are coupled through quantum interconnects to form a larger processor, has been suggested as an extensible solution~\cite{Vandersypen17}. 
Such an architecture requires entangling gates between the qubits connected by these interconnects.
Superconducting circuits are reasonable choices for interconnects for two primary reasons.
First, their larger length scale (often of order $\SI{100}{\micro m}$ or more) can increase the separation between quantum dots and thereby ease wiring density.
Second, they can couple to electric dipoles, which are readily induced in quantum dots~\cite{Mi17, Burkard20}.

Previous research has concentrated on superconducting resonators as the interconnects.
In such architectures, spin-photon interactions couple two distant quantum dots to either end of a microwave resonator, leading to an effective spin-spin coupling between the two spin qubits~\cite{Imamog99, Childress04}.
Such interactions have been proposed and demonstrated in double-quantum-dot (DQD)~\cite{Trifunovic12, Mi17, Woerkom18, Pomorski19, Borjans20, Collard22, Dijkema25}, singlet-triplet (ST)~\cite{Harvey18, Bottcher22, Ungerer24}, and resonant-exchange (RX) qubits~\cite{Landig18,Landig19,Scarlino19}.  These types of resonator-mediated coupling concepts are also applied to other systems, e.g., nitrogen-vacancy (NV) centers coupled to mechanical oscillators, to give just one example~\cite{Rabl10, Rosenfeld21}.
However, materials and fabrication limitations---for example, the achievable impedance of high kinetic-inductance resonators~\cite{Blais04, Harvey18}---limit the resulting coupling strength. 
Thus, reported coupling strengths have been generally comparable to decoherence rates, preventing a high-fidelity two-qubit gate.

One way to bypass this limitation is to use a different superconducting circuit as the quantum interconnect. In superconducting-qubit architectures,  nonlinear lumped-element circuits have been employed as couplers between qubits~\cite{Niskanen06, Chen14, Fei18, Sung21, Simakov23, Ding23}.
The coupler circuit should be chosen based on the type of qubit.
In particular, for spin qubits with induced electric dipoles, it is natural to select a charge-sensitive circuit. One such circuit is the offset-charge-sensitive (OCS) transmon, which has been used as a charge sensor for investigating quasiparticle tunneling events~\cite{Ristè13, Serniak19}. Its charge sensitivity can be adjusted to the application, ranging from the highly charge-sensitive Cooper-pair box (CPB) regime~\cite{Bouchiat98, Nakamura99} to the transmon regime~\cite{Koch07} of suppressed charge sensitivity.

\begin{figure*}[ht!]
\subfloat{\label{fig:system_schematic}}
\subfloat{\label{fig:rx_qubit}}
\subfloat{\label{fig:rx_electron_number}}
\subfloat{\label{fig:frequency_shift}}
\subfloat{\label{fig:drive_principle}}
\includegraphics{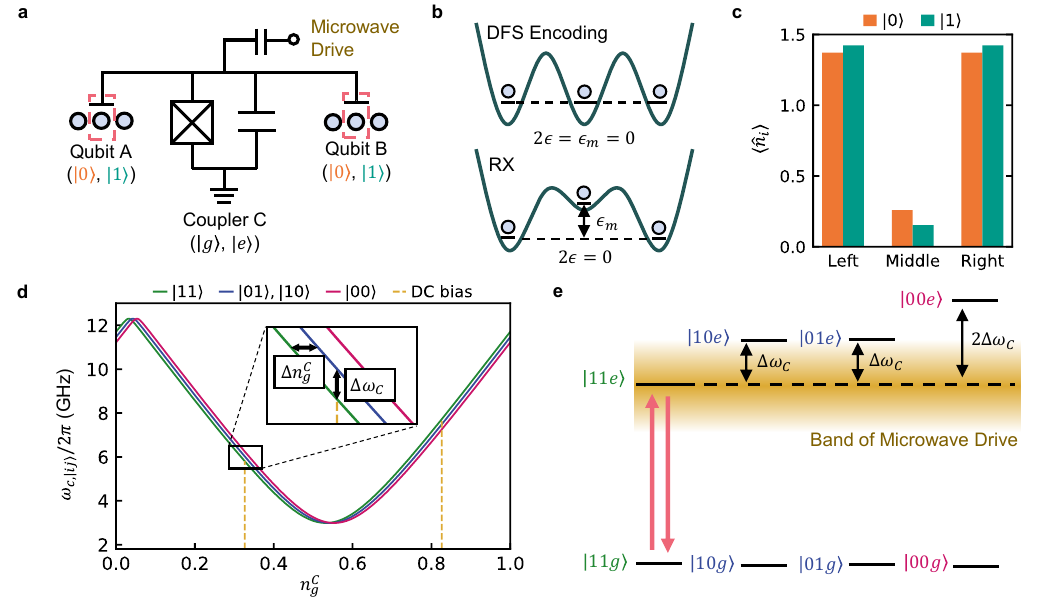}
\caption{\textbf{System overview}. \textbf{(a)}~Schematic of the hybrid system. The OCS transmon coupler is capacitively coupled to the middle dot of each qubit. A detailed circuit diagram is shown in Appendix~\ref{sec:a_quantization}. \textbf{(b)}~Diagram of the potential landscape of an DFS-encoded EO qubit and an RX qubit. The middle dot is energetically unfavorable for electrons by a difference of $\epsilon_m$ compared to the left and right dots. \textbf{(c)}~Expected number of electrons occupying each dot, shown for the $|0\rangle^D$ and $|1\rangle^D$ RX qubit states. The logic states correspond to different occupancy of the middle dot. Fermi-Hubbard parameters of $U_C=0.2U$, $\epsilon_m = 0.625U$, and $t=0.013U$ are used. 
\textbf{(d)}~Qubit-state-dependent coupler frequency shift. The difference in effective gate charge $\Delta n^C_g$ leads to a difference in coupler frequency, $\Delta \omega_c$. For identical qubits A and B, $|01\rangle$ and $|10\rangle$ states have the same energy. Here, we use $E_J/h=3$~GHz, $E_C/h = 3$~GHz, and $\alpha^D=0.2$. The example bias voltages $n^C_{g}$ are marked with yellow dashed line, separated by 0.5 due to possible quasiparticle tunneling, as explained in Sec.~\ref{sec:3.3}.
\textbf{(e)}~ Energy-level diagram (not to scale). Note that the carrier frequency and bandwidth of the drive dependends on the specific pulse shape. For simplicity, ground state energy level shifts are neglected, since they do not affect CPhase$(\theta)$ gates as explained in Appendix~\ref{sec:a_drive_matrix_representation}.
} 
\label{fig:fig1}  
\end{figure*}

In this work, we propose to use an OCS transmon coupler to mediate an interaction between two distant three-electron spin qubits biased in the charge-sensitive RX regime~\cite{medford13}. The coupler is capacitively coupled to the center dot of both RX qubits, as shown in Fig.~\ref{fig:system_schematic}. This leads to a \textit{longitudinal} interaction, unlike previous works that focused on \textit{transverse} interactions between an RX qubit and a transmon mediated by a resonator~\cite{Landig18,Landig19,Scarlino19}.
For a longitudinal interaction, the qubit states determine the coupler frequency. Therefore, a pulse applied to the coupler will result in a rotation that is conditional on the states of the qubits. In this way, we implement a conditional phase shift on the overall wavefunction, i.e., a CPhase($\theta$) gate.

This paper is organized as follows.
In Section~\ref{sec:2}, we derive the qubit logical state-dependent coupler gate-charge shift $\Delta n^C_g$ and frequency shift $\Delta \omega_c$, based on a capacitance network model of the hybrid quantum dot-superconducting circuit system. This model generalizes the field-dipole model employed in earlier works~\cite{Blais04, Srinivasa16, Russ17, Harvey18, Burkard20}.
In Section~\ref{sec:3}, we discuss strategies to mitigate noise sources and provide two implementations of a CZ gate, one without and one with dynamical decoupling. 
In the former, an off-resonant square pulse provides a fast gate.
In the latter, we propose to surround a dynamical decoupling pulse by two $\sqrt{\mathrm{CZ}}$ gates, which are realized using pulse-envelope engineering, to mitigate the low-frequency noise prevalent in solid state systems.
In Section~\ref{sec:4}, we suggest optimal quantum dots and superconducting circuits parameter and calculate the expected gate fidelities based on realistic noise models.
We consider the two dominant noise sources: $1/f^\beta$ charge noise on the spin qubits and on the coupler, where $\beta \approx 1$~\cite{Astafiev04, Bylander11, Dial13, Gustafsson13, Elliot22, Zwerver22}. 
We determine the infidelity from the spin-qubit dephasing analytically. 
On the other hand, the infidelity from the coupler dephasing is numerically simulated in the time domain, based on the 2$^{\mathrm{nd}}$-order truncation of the generalized cumulant expansion, to accurately model the dephasing process for the driven system~\cite{Kampen74, Groszkowski23}.
The simulation results indicate gate fidelities exceeding 90\% are possible when including $1/f$-like charge noise models.
Interestingly, we find that the pulse sequence that provides superior performance depends on the charge noise profiles of the spin qubits and of the coupler.

\section{\label{sec:2} System and Model}
Our goal is to actuate a remote entangling gate between two exchange-only (EO) spin qubits hosted in gate-defined quantum dots (Fig.~\ref{fig:system_schematic}).
Each EO qubit encodes its logical state in the spin degree of freedom of a three-electron wavefunction inside of three quantum dots. When the qubits idle or undergo single-qubit gate operations, the dots are set to equal chemical potentials (Fig.~\ref{fig:rx_qubit}). 
The logical states then have equal spin projections and charge distributions and correspondingly offer protection from global magnetic and charge fluctuations~\cite{Russ17}.
Storing information in these states is therefore referred to as the decoherence-free subspace (DFS) encoding.

Being spin qubits, EO qubits naturally interact via electron-exchange interactions, offering a straightforward mechanism for two-qubit gates between adjacent qubits~\cite{Weinstein23}. 
However, the short range of exchange interactions precludes long-range coupling based on the spin degree of freedom.
Instead, long-range coupling is more easily realized via the charge degree of freedom.
Such capacitive coupling requires that the logical states have different charge distributions, which is inconsistent with a DFS encoding.
During long-range two-qubit gate operations, we therefore adiabatically transition both qubits to the resonant-exchange (RX) regime, where the chemical potential of each qubit's center dot is raised (Fig.~\ref{fig:rx_qubit}).
The logical states in the RX regime share the spin quantum numbers---total spin $S = 1/2$ and total spin component along the quantization axis $S_z = 1/2$ or $S_z = -1/2$, or any mixture of the two~\cite{Laird10}---with their DFS-encoded counterparts.
Yet, because the center dot is energetically unfavorable, in the RX regime the charge distribution is shifted towards the outer dots.
Crucially, the two logical states have distinct charge distributions due to Pauli spin blockade~\cite{Burkard23}, as shown in Fig.~\ref{fig:rx_electron_number}.
This property of the RX regime, called spin-charge conversion, allows spin qubits to interact with a long-range coupling element via electric fields.

We use an OCS transmon as the long-range coupling element mediating interactions between the two qubits.
An OCS transmon has a circuit structure identical to that of CPBs and conventional transmons. The sensitivity of energy levels to external charge in such circuit structures is determined by the ratio between the inductive energy of the JJ ($E_J$) and the single-electron charging energy of the capacitor ($E_C$). Notably, an OCS transmon features an intermediate capacitance---smaller than that of a transmon ($E_J/E_C \gtrsim 50$)~\cite{Koch07}, but larger than that of a CPB ($E_J/E_C \sim 0.1$)~\cite{Bouchiat98, Nakamura99}.
Its energy levels are therefore sensitive to external charge, but not so sensitive as to be overwhelmed by charge noise.
Thus, when capactively coupled to two RX qubits, the energy levels of the OCS transmon coupler are conditioned by logical states of both qubits. 

\subsection{\label{sec:2.1} Interaction Hamiltonian}

We derive the form of the quantum dot-OCS transmon interaction using a capacitive-network model.
We describe the undriven system using the Hamiltonian
\begin{eqnarray}\label{eq:H0}
    \hat{H}_0~&=&\sum_{D=A,B} \hat{H}^D + \hat{H}^C + \hat{H}_\mathrm{int},
\end{eqnarray}
where $\hat{H}^D$ is the Hamiltonian of each spin qubit $(D=A,\, B)$, $\hat{H}^C$ is the Hamiltonian of the OCS transmon coupler, and $\hat{H}_\mathrm{int}$ describes interactions between the OCS transmon and each spin qubit.
A time-dependent drive responsible for rotating the state of the coupler will be introduced later (Section~\ref{sec:2.3}).

We describe the spin qubits using a Fermi-Hubbard model~\cite{Russ17}: 
\begin{eqnarray}
    \hat{H}^D & = &\frac{U^D}{2} \sum_{i} \hat{n}^D_{i} (\hat{n}^D_{i}-1) + U^D_C \sum_{\langle i,j \rangle} \hat{n}^D_{i} \hat{n}^D_{j} \nonumber \\
    &&+ \sum_{i} V_i^D\hat{n}^D_{i} - \sum_{\langle i,j \rangle, \sigma} t^D_{ij}\left( \hat{c}^{\dagger D}_{i\sigma} \hat{c}^D_{j\sigma} + \text{H.c.} \right),\label{eq:fh}
\end{eqnarray}
where $i=1,2,3$ indexes the three dots comprising each spin qubit, $\hat{n}^D_i=\hat{Q}^D_i/e$ is the total electron number operator for site $i$, and $\hat{c}^D_{i\sigma}$ is the corresponding fermionic annihilation operator for an electron with spin $\sigma$. 
The first term describes on-site Hubbard interactions with strength $U^D$. 
The second term describes intersite Hubbard interactions, which we approximate as being nearest-neighbor-only, with strength $U_C^D$. 
The third term describes the on-site energies $V_i^D$, which are set by the gate voltages.
The final term represents particle exchange between neighboring lattice sites with amplitude $t^D_{ij}$.
In this work, we assume $t^D_{12}=t^D_{23}=t$ for simplicity.
The qubit operation regime is determined by the detuning parameters $\epsilon^D_m = V^D_2 - (V^D_1 + V^D_3)/2 + U^D_C$ and $\epsilon^D = (V^D_1 - V^D_3)/2$.
For EO qubits, $\epsilon^D=0$.
For the DFS encoding, $\epsilon^D_m$ is also zero, whereas for the RX regime, $\epsilon^D_m>0$, and larger values correspond to larger charge dipoles.
Further details on the Fermi-Hubbard model are provided in Appendix~\ref{sec:a_rx}.

The OCS transmon Hamiltonian $\hat{H}^C$ in the first-quantized form is~\cite{Krantz19}
\begin{equation}
    \hat{H}^C=4E_C(\hat{n}^C-n^C_g)^2-E_J\cos(\hat{\phi}^C), \label{eq:transmon_h}
\end{equation}
where $\hat{n}^C=\hat{Q}^C/(2e)$ is the Cooper-pair number operator, $n^C_g$ is the gate charge determined by an applied gate voltage, and $\hat{\phi}^C$ is the phase across the Josephson junction. The superscript $C$ indicates that the operators act on the coupler state.
Note that we use the single electron (units of $e$) and the superconducting (units of $2e$) convention for spin qubit and coupler degrees of freedom, respectively.
The inductive energy of the transmon is $E_J$, and its capacitive energy is $E_C=e^2/(2C^C_{\Sigma})$, where the total capacitance $C^C_{\Sigma}$ includes the coupling capacitance to the quantum dots, which may be non-negligible.

The interaction Hamiltonian can be found by quantizing the entire system as a whole. This involves treating quantum dots as a capacitive network of metallic islands~\cite{Wiel02} and extracting the terms that involve both quantum dot and coupler operators. When the capacitance to ground dominates the total capacitance for each node, the interaction Hamiltonian approximated to first order is
\begin{eqnarray}
    \hat{H}_{\textrm{int}} &\simeq& \sum_{D=A,B} \sum_{i=1,2,3} 4 E_C \alpha_i^D \hat{n}^D_i \hat{n}^C, \label{eq:intH_2}
\end{eqnarray}
where $\alpha^D_i$ is the lever arm between the coupler and the $i^{\mathrm{th}}$ dot.
The above is derived in Appendix~\ref{sec:a_quantization}.

The above expression generalizes the field-dipole interaction model used in several earlier works~\cite{Srinivasa16, Russ17, Burkard20}, which describes the electric dipole of the quantum dots interacting with the electric field of a resonator.
The field-dipole interaction model is: 
\begin{equation}
    \hat{H}_\mathrm{field-dipole} = \sum_{D=A,B} i g \hat{\sigma}^D_k (\hat{a}^R - \hat{a}^{\dagger R}), \label{eq:field_dipole}
\end{equation}
where $g$ is the coupling strength, $\hat{a}^R$ is the second-quantized annihilation operator for such a resonator, and $\hat{\sigma}^D_k$ are qubit Pauli operators. 
With respect to the qubit, the field-dipole interaction model describes a transverse interaction for $k\rightarrow x$, and it describes a longitudinal interaction for $k\rightarrow z$. 
With respect to the resonator, the interaction is transverse (though appears longitudinal in the dispersive limit~\cite{Krantz19}).

In contrast, our model describes the interaction with first-quantized electron number operators $\hat{n}^D_i$ and $\hat{n}^C$. 
The relationship between Eq.~\ref{eq:intH_2} and~\ref{eq:field_dipole} may be seen by defining ladder operators for the coupler $\hat{a}^C$ such that $\hat{n}^C =n^C_{\text{ZPF}}\times i(\hat{a}^C - \hat{a}^{\dagger C})$, where $n^C_{\text{ZPF}}=(E_J/32E_C)^{1/4}$ is the zero-point fluctuation of the coupler charge~\cite{Krantz19}, and by decomposing $\hat{n}_i^D$ into Pauli matrices in the qubit logical subspace, which we will do momentarily.

We focus on the interaction model Eq.~\ref{eq:intH_2} for two reasons.
First, the coupler modes are not eigenstates of $\hat{a}^{\dag{} C} \hat{a}^C$, unlike the case of resonators. 
Therefore, $\hat{a}^C|N\rangle \neq \sqrt{N} |N-1\rangle$ for coupler modes $|N\rangle$. 
As a consequence, here $\hat{a}^C - \hat{a}^{\dagger C}$ is neither purely transverse nor purely longitudinal, but rather includes components of both with strengths depending on $n^C_g$.
Our gate proposal relies on the longitudinal component of this operator, and we use the notation $\hat{n}^C$ to avoid confusion (as $\hat{a}^R - \hat{a}^{\dagger R}$ is purely transverse for resonators).
Secondly, working in terms of $\hat{n}^C$ is convenient because, as we will see shortly, the logical states of the spin qubits act as a gate-charge shift on the coupler.

We are interested in the case where the interaction reduces to a purely longitudinal coupling:
\begin{eqnarray}
    \hat{H}_{\textrm{int}}=\sum_{D=A,B}\left(g^D\hat{\sigma}_z^D + g^D_I \hat{I} \right)\frac{\hat{n}^C}{n^C_\text{ZPF}},
    \label{eq:intH_3}
\end{eqnarray}
where $g^D$ is the coupling strength corresponding to qubit Pauli Z operator $\hat{\sigma}_z^D$ and $g^D_I$ is the qubit-state-independent coupling strength.
Purely longitudinal coupling is achieved when $\alpha^D_1 = \alpha^D_3$, i.e. when the left and the right qubits are coupled to the OCS transmon identically.

Under this condition, Eqs.~\ref{eq:intH_2} and~\ref{eq:intH_3} yield
\begin{eqnarray}
    g^D&=&2E_C n^C_\text{ZPF}\{\alpha^D \left(\langle0|\hat{n}_2|0\rangle^D-\langle1|\hat{n}_2|1\rangle^D\right)\},\label{eq:g_d}\\ 
    g^D_I &=& 2E_C n^C_\text{ZPF} \{\alpha^D \left(\langle0|\hat{n}_2|0\rangle^D +\langle1|\hat{n}_2|1\rangle^D \right) + 6\alpha_1^D\} \label{eq:g_I}
\end{eqnarray}
where $\alpha^D = \alpha^D_2-\alpha^D_1 = \alpha^D_2 - \alpha^D_3$ is the differential lever arm of the qubit, and $\langle 0|\hat{n}^D_2| 0\rangle^D$, $\langle 1|\hat{n}^D_2| 1\rangle^D$ are the expectation values of the number of electrons in the middle quantum dot, for single-qubit states $|0\rangle^D$ and $|1\rangle^D$, respectively.
The derivation of the above expressions is shown in Appendix~\ref{sec:a_longitudinal}.
As shown in Fig.~\ref{fig:rx_electron_number}, the number of electrons in the middle dot differs between the logical RX states, leading to the longitudinal interaction. In the DFS encoding, where $\epsilon=\epsilon_m=0$ and $\langle0|\hat{n}^D_2|0\rangle^D=\langle1|\hat{n}^D_2|1\rangle^D=1$, the qubit-state-dependent interaction vanishes.

\subsection{\label{sec:2.2} Qubit-State Dependent \\ Coupler Frequency Shift}

The interaction Hamiltonian Eq.~\ref{eq:intH_3} can be interpreted as a qubit-logic-state-dependent shift of the coupler gate charge. This can be seen by combining $\hat{H}^C$ and $\hat{H}_{\textrm{int}}$:
\begin{eqnarray}
    \hat{H}^C+\hat{H}_{\textrm{int}}&=&4E_C\left(\hat{n}^C-n^C_g+\sum_{D=A,B}\frac{g^D\hat{\sigma}_z^D + g^D_I \hat{I}}{8E_C n^C_\text{ZPF}}\right)^2 \nonumber\\&&-E_J\cos(\hat{\phi}^C) + f(n^C_g, \hat{\sigma}_z^D),
    \label{eq:H_combined}
\end{eqnarray}
where $f(n^C_g, \hat{\sigma}_z^D)$ is the term independent of $\hat{n}^C$. Equation~\ref{eq:H_combined} includes terms up to second order in the coupling strength $g^D$; higher orders are negligible for realistic coupling strengths. Thus, the OCS transmon experiences a qubit-state dependent shift in the gate charge:
\begin{equation}
    \hat{n}^C_{g,\mathrm{eff}} = n^C_g-\sum_{D=A,B}\frac{g^D\hat{\sigma}_z^D + g^D_I \hat{I}}{8E_C n^C_\text{ZPF}}\;.
\end{equation} 

In this proposal, we take
the qubits to be coupled to the OCS transmon identically for easier control: $g^A=g^B$. As a result, $\langle \hat{n}^C_{g,\mathrm{eff}}\rangle$ becomes the same for the $|01\rangle$ and $|10\rangle$ qubit states. Moreover, $\langle \hat{n}^C_{g,\mathrm{eff}}\rangle$ of $|00\rangle$, $|01\rangle$, and $|11\rangle$ are evenly spaced. We define this spacing between the gate-charge shift between the logical states as $\Delta n^C_g$:
\begin{eqnarray}
    \Delta n^C_g &= & \left|\frac{g^D}{4E_C n^C_\text{ZPF}}\right| =\frac{\alpha^D
    |\langle0|\hat{n}_2|0\rangle^D-\langle1|\hat{n}_2|1\rangle^D|}{2} .\label{eq:del_n_g_fh}
\end{eqnarray}

As mentioned earlier, the energy levels of OCS transmons are dependent on gate charge. Thus, the shift in effective gate charge is translated into the shift in coupler frequency $\omega_c$. 
The coupler frequency in units of GHz for the example circuit and quantum dot parameters are plotted in the charge dispersion diagram Fig.~\ref{fig:frequency_shift}.
In the linear regime where the slope is nearly constant, the coupler frequencies conditioned by each 2-qubit state are also equally spaced, for that specific controlled DC bias $n^C_{g,0}$ is equal. Hence, the frequency shift in the linear regime $\Delta \omega_c=|\omega_{c,|11\rangle}-\omega_{c,|10\rangle}|\simeq|\omega_{c,|01\rangle}-\omega_{c,|00\rangle}|$ is
\begin{eqnarray}
    \Delta \omega_{c}&\simeq&\left|\frac{\partial \omega_{c}}{\partial n^C_g}\right|\Delta n^C_g
\end{eqnarray}
where $\omega_{c,|ab\rangle}$ is the coupler frequency conditioned by the 2-qubit state $|ab\rangle$. Note that in the limit of $E_J/E_C \rightarrow 0$, $|\partial \omega_{c}/\partial n^C_g| \rightarrow 4E_C$, which can be shown by differentiating Eq.~\ref{eq:transmon_h} with respect to $n^C_g$. While the OCS transmon coupler has $E_J \geq E_C$, this limit still provides a rough estimate of the typical scale of the frequency shift, deviating by 8\% for the optimal coupler chosen in Sec.~\ref{sec:4}.

\subsection{\label{sec:2.3}Gate Operation Principle}
The essence of the microwave-activated phase gate is to drive the coupler through a closed unitary trajectory that depends on the logical qubit states, while keeping the qubits tuned to the RX regime. 
This is possible because the transition frequency between the coupler ground and excited states depends on the qubit states, as seen in Fig.~\ref{fig:drive_principle}.
As the phase acquired during the evolution for a given qubit state depends on the solid angle enclosed by the trajectory on the coupler Bloch sphere~\cite{Li20}, a difference in the trajectory leads to a qubit-state-dependent conditional phase.

We apply a microwave drive through the coupler gate charge, in addition to the constant bias $\hat{n}^C_{g,0}$. This leads to a transverse drive Hamiltonian $\hat{H}_{\mathrm{drive}}(t)$:
\begin{eqnarray}
    \hat{H}_{\mathrm{drive}}(t) = \hbar \left(\Omega_x(t)\cos(\omega_d t)  + \Omega_y(t)\sin(\omega_d t)\right) \hat{\sigma}^C_x, \label{eq:drive_H_lab}
\end{eqnarray}
where $\Omega_x(t)$ and $\Omega_y(t)$ are the slowly varying in-phase and out-of-phase envelopes, $\omega_d$ is the carrier frequency, and $\hat{\sigma}^C_x$ is the coupler Pauli-X operator.
Note that perturbing the gate charge would also induce longitudinal drive terms, as the coupler charge operator $\hat{n}^C$ would not be purely transverse in general, unlike for conventional transmons. However, the longitudinal drive terms can be ignored within the rotating wave approximation. The full Hamiltonian of the driven system after the rotating wave approximation, $\hat{H}_R$, in matrix notation is explicitly shown in Appendix~\ref{sec:a_drive_matrix_representation}. 

\begin{figure*}[ht!]
\subfloat{\label{fig:non_DD_sequence}}
\subfloat{\label{fig:DD_sequence}}
\includegraphics{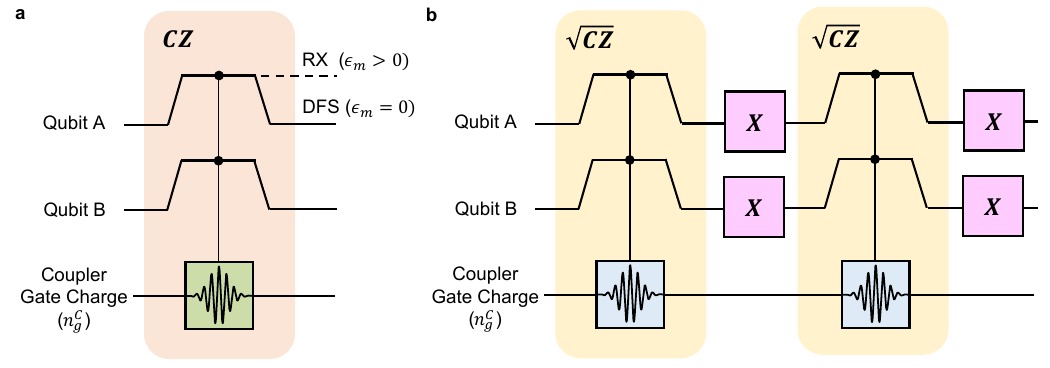}
\caption{\textbf{CZ gate pulse sequence with and without dynamical decoupling}. \textbf{(a)}~Pulse sequence for a CZ gate without Ddynamical decoupling. When the coupler is microwave-driven, the qubits are biased to charge-senstive RX regime by increasing the middle dot potential $\epsilon_m$. Otherwise, they are set in DFS encoding where $\epsilon_m=0$. \textbf{(b)}~Pulse sequence for a CZ gate with dynamical decoupling. Alternating $\sqrt{\mathrm{CZ}}$ gates and single qubit $X$ gates twice lead to CZ gate, which is proven in Appendix~\ref{sec:a_DD}. With this sequence, dephasing from low-frequency charge noise on qubits is echoed out.
} 
\label{fig:fig3}  
\end{figure*}

The goal of a CPhase($\theta$) gate is to accumulate a phase $\theta$ only for the $|11\rangle$ qubit state. After the unitary gate evolution, each coupler subspace spanned by $|ab\rangle \otimes|g\rangle$ and $|ab\rangle \otimes |e\rangle$, for 2-qubit states $|ab\rangle$, accumulates a phase $\theta_{ab}$, where we use $|g\rangle$ and $|e\rangle$ to represent the ground state and first excited state of the coupler, respectively, to distinguish them from the 0 and 1 used for the qubit states.
A na\"ive approach is to drive resonantly with the coupler transition conditioned on $|11\rangle$ to give $\theta_{11}=\theta$ and $\theta_{00} = \theta_{01} = \theta_{10}$ = 0.
However, because $\Delta \omega_c$ is small, such a gate would need to be exceptionally slow to avoid off-resonantly driving the coupler transitions aside from the $|11\rangle$ condition.
Instead, we allow for single-qubit phase corrections such that the condition becomes more lenient~\cite{Krantz19}:
\begin{equation}
    \theta = (\theta_{11} - \theta_{10}) - (\theta_{01} - \theta_{00}). \label{eq:central_phase_eq}
\end{equation}
The design of a CPhase($\theta$) gate is thus choosing appropriate $\omega_d$, $\Omega_x(t)$, and $\Omega_y(t)$, so that after unitary evolution $\theta_{ab}$ satisfies Eq.~\ref{eq:central_phase_eq}.
Following the conditional phase gate, single-qubit gates are then used to null the phases $\theta_{00}$,  $\theta_{01}$, and $\theta_{10}$.

\section{\label{sec:3} Noise Mitigation}

The gate is susceptible to incoherent dephasing errors due to charge noise on the coupler and on the qubits, which typically assumes a spectral density scaling with frequency $f$ as $1/f^\beta$ with $0.6<\beta<1.4$~\cite{Astafiev04, Bylander11, Dial13, Gustafsson13, Harvey18, Elliot22, Zwerver22}.
Other decoherence mechanisms, including dephasing from hyperfine magnetic noise~\cite{Petta05} on quantum dots or microwave photon loss~\cite{Mi17} on the OCS transmon, contribute $\sim10^{-2}$ infidelity for the typical timescales (see Sec.~\ref{sec:3.4}).

Here, we present two types of gates to mitigate dephasing errors.
First, we discuss a rapid CZ gate with an off-resonant drive, inspired by recent work with superconducting qubits~\cite{Simakov23} albeit incompatible with dynamical decoupling. 
Second, we develop a CZ gate that suppresses the low-frequency portion of qubit-dephasing noise through a dynamical decoupling sequence.
The dynamically decoupled CZ gate comprises two $\sqrt{\mathrm{CZ}}$ gates with the qubits in the RX regime, after each of which single-qubit $X$ gates are executed with the qubits within the DFS. 
This sequence is related to a prior proposal for decoupled CZ gates~\cite{Watson18, Huang24}. 
The $\sqrt{\mathrm{CZ}}$ gates are realized through a systematic pulse shaping approach to enable appropriate phase acquisition for all four conditional coupler transitions (whereas square and Gaussian pulses, for example, do not). 
Our approach generalizes to CPhase($\theta$) gates with arbitrary acquired phase $\theta$ and is resilient to quasiparticle tunneling events across the coupler Josephson junction.
Pulse sequences for a CZ gate with and without dynamical decoupling are compared in Fig.~\ref{fig:fig3}.

\subsection{\label{sec:3.1}Constant-Amplitude, Off-Resonant CZ Gate}
\begin{figure}[ht!]
\subfloat{\label{fig:off_cz_sig_z_trajectory}}
\subfloat{\label{fig:off_cz_bloch}}
\includegraphics{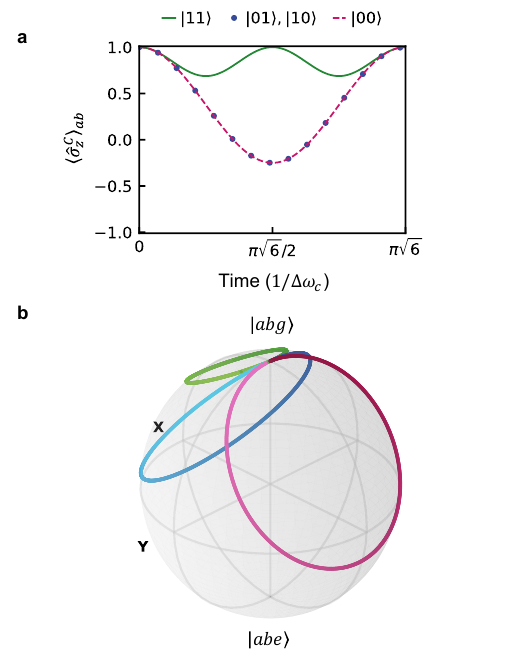}
\caption{\textbf{Off-resonant CZ gate with constant drive amplitude}. \textbf{(a)}~Evolution of coupler excited state population throughout the gate, $\langle \hat{\sigma}^C_z\rangle_{ab}$. Only qubit $|11\rangle$ state leads to oscillation twice, while other states induce single oscillation. As the drive is off-resonant, a full population inversion is not achieved. \textbf{(b)}~Trajectory of the states on the coupler Bloch sphere. The dark and bright colors indicate the start and end of the evolution, respectively. The smaller area enclosed by the coupler trajectory conditioned by $|11\rangle$ state leads to a smaller phase acquisition. For none of the qubit states does the coupler undergo equatorial rotation. 
} 
\label{fig:fig2}  
\end{figure}
One approach for a fast CZ gate is to drive off-resonance from all coupler transitions so that the coupler accumulates a different phase for different qubit logical states.
When $\omega_{c,|11\rangle}<\omega_{c,|10\rangle}$ (the bias point shown in the inset of Fig.~\ref{fig:frequency_shift}), a CZ gate follows from the conditions $\omega_d = \omega_{c,|11\rangle} + 3\Delta \omega_c/2$, $\Omega_x = \sqrt{5/12} \Delta \omega_c$, $\Omega_y = 0$~\cite{Simakov23}, and gate time
\begin{eqnarray}
    t_g&=&\frac{2\pi}{\sqrt{\Omega^2+(\Delta \omega_c/2)^2}}=\frac{\pi\sqrt{6}}{\Delta \omega_c}.
\end{eqnarray}
When $\omega_{c,|11\rangle}>\omega_{c,|10\rangle}$, the condition for the drive frequency becomes $\omega_d = \omega_{c,|11\rangle} - 3\Delta \omega_c/2$.
The trajectories of the coupler state conditioned on each qubit logical state are shown in Fig.~\ref{fig:fig2}; an additional $\pi$ phase is accumulated for the $|11\rangle$ logical state. Under the rotating wave approximation and a perfectly identical coupling of the two qubits, the gate applies a $CZ$ unitary with zero coherent error.

\subsection{\label{sec:3.2}Pulse Envelope Engineering}
\begin{figure*}[ht!]
\subfloat{\label{fig:sqrt_cz_pulse_envelope}}
\subfloat{\label{fig:sqrt_cz_pulse_envelope_freq}}
\subfloat{\label{fig:sqrt_cz_sig_z_trajectory}}
\subfloat{\label{fig:sqrt_cz_bloch}}
\includegraphics{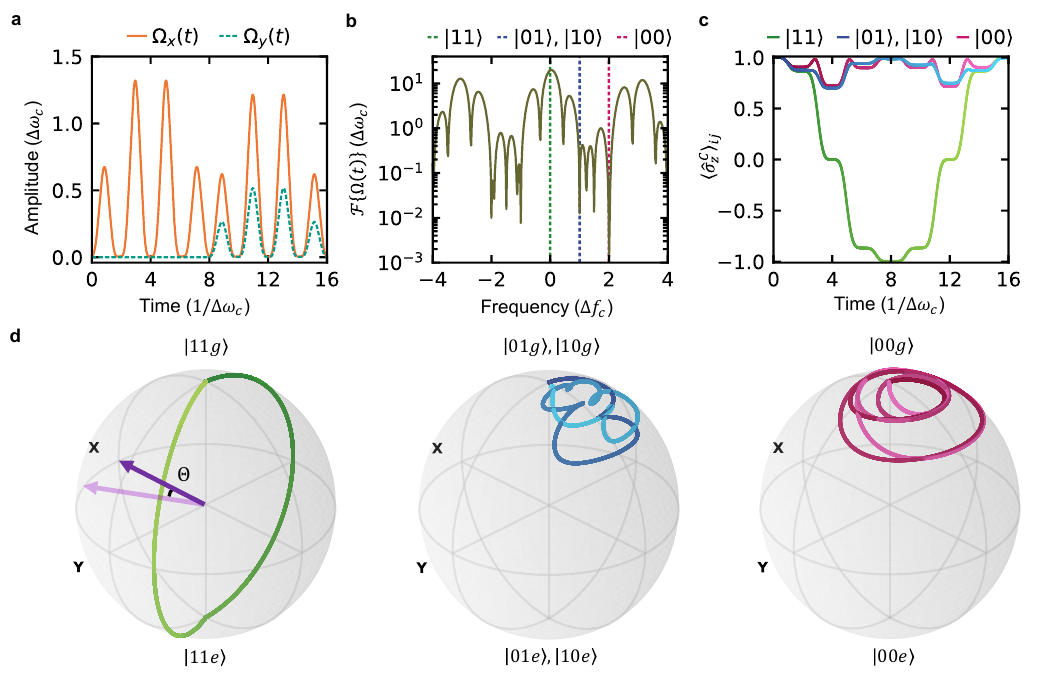}
\caption{\textbf{$\sqrt{\mathrm{CZ}}$ gate based on pulse envelope engineering}. \textbf{(a)}~Pulse envelope in time domain. The second half of the pulse in the time domain is the rotated version of the first half. \textbf{(b)}~Pulse envelope in frequency domain. The pulse has negligible components near the transition frequencies conditioned by $|01\rangle$, $|10\rangle$, and $|00\rangle$ qubit states. Multiple sidelobes occur from cropping of the pulse, due to finite gate time. \textbf{(c)}~Evolution of excited state coupler population throughout the gate. The dark and bright colors indicate the start and end of the evolution. After the first half of the drive, the full population inversion of the coupler is achieved when conditioned by qubit $|11\rangle$ state. Otherwise, the coupler returns to the ground state. \textbf{(d)}~Trajectory of the states on each coupler Bloch sphere. The dark and bright colors indicate the start and end of the evolution. The coupler evolves along the meridian when conditioned by $|11\rangle$ qubit state, as the drive is resonant for that subspace. Both the trajectory and the drive axis (purple arrow) are tilted after the first half by angle $\Theta$. This is to adjust the solid angle of the area enclosed by the trajectory, which has a linear relationship with the conditional phase~\cite{Li20}.
} 
\label{fig:fig4}  
\end{figure*}

The CZ gate in Sec.~\ref{sec:3.1} has the advantage that it is relatively fast, but it also features a fixed phase acquisition $\theta=\pi$ and thus is not suitable for the types of pulses (e.g., $\theta=\pi/2$) needed for dynamical decoupling. In this subsection, we present a generalized approach to implement a fast CPhase$(\theta)$ gate with arbitrary $\theta$ and the resulting $\sqrt{\mathrm{CZ}}$ gate. The general solution we found is slower than the constant-amplitude, off-resonant CZ gate of Sec.~\ref{sec:3.1}, but provides adjustable conditional phase $\theta$ and therefore permits a dynamical decoupling sequence.

The idea is to divide the gate into two stages, where the drive at the second stage differs by a phase $\Theta$ compared to the drive at the first stage. Fig.~\ref{fig:fig4} depicts a $\sqrt{\mathrm{CZ}}$ gate pulse found through this method. In the first stage, the drive is applied to fully excite the coupler only when conditioned by $|11\rangle$ qubit state (Fig.~\ref{fig:sqrt_cz_sig_z_trajectory}). Otherwise, the coupler returns to the ground state. In the second stage, we apply the same drive, but shifted by a phase $\Theta$. On the Bloch sphere, this translates into the rotation of the drive axis by the azimuthal angle $\Theta$, as shown in Fig.~\ref{fig:sqrt_cz_bloch}. The solid angle enclosed by the green trajectory on the Bloch sphere of $|11g\rangle$ and $|11e\rangle$ state is proportional to $\Theta$. Thus, $\Theta$ can be chosen to compensate for the phase acquired by the trajectories that arise for the other qubit states $|01\rangle$, $|10\rangle$, and $|11\rangle$, leading to CPhase($\theta$) gates with adjustable $\theta$. For the purpose of dynamically decoupled CZ gate, we aim for $\theta=\pi/2$.

Quantitatively, the above method requires finding the $\Omega_0(t)$, $t_g$, and $\Theta$ that satisfy the constraints:
\begin{eqnarray}\label{eq:omega_x_condition}
    \Omega_x(t) = \begin{cases}
        \Omega_0(t) &(t<\frac{t_g}{2}) \\
        \Omega_0(t - \frac{t_g}{2})\cos{\Theta} &(\frac{t_g}{2}<t<t_g)
    \end{cases}, \label{eq:omega_x}\\
    \Omega_y(t) = \begin{cases}
        0 &(t<\frac{t_g}{2}) \\
        \Omega_0(t - \frac{t_g}{2})\sin{\Theta} &(\frac{t_g}{2}<t<t_g)
    \end{cases} \label{eq:omega_y},
\end{eqnarray}
where at $t=\frac{t_g}{2}$:
\begin{eqnarray}
    \langle \hat{\sigma}_z^C \rangle_{ab}\left(t=\frac{t_g}{2}\right) = \begin{cases}
        1 &(ab = 00, 01, 10) \\
        -1 &(ab = 11)
    \end{cases},
    \label{eq:gamam_cond_1}
\end{eqnarray}
and
\begin{eqnarray}
       \theta\left(\hat{H}_R\left(\omega_d, \Omega_0(t)\right), t=t_g\right) = \frac{\pi}{2}
       \label{eq:gamam_cond_2}
\end{eqnarray}
at the end of the gate. $\hat{H}_R$ is the Hamiltonian of the driven system, explicitly shown in Eq~\ref{eq:driven_H_RWA} of Appendix~\ref{sec:a_drive_matrix_representation}.

Equations \ref{eq:omega_x_condition}-\ref{eq:gamam_cond_2} may have multiple solutions. In this work, we focus on resonant gates, where the drive is resonant with the coupler transition for the $|11\rangle$ logical state ($\omega_d = \omega_{c,|11\rangle}$). For a resonant drive, the condition Eq.~\ref{eq:gamam_cond_1} is approximated to
\begin{eqnarray}
    &&\int_0^{\frac{t_g}{2}} dt \Omega_0(t) = \pi \label{eq:gamam_cond_3},\\
    &&\int_0^{\frac{t_g}{2}} dt \Omega_0(t)e^{-i\Delta\omega_ct} = 0
    \label{eq:gamam_cond_4},\\
    &&\int_0^{\frac{t_g}{2}} dt \Omega_0(t)e^{-i2\Delta\omega_ct} = 0,
    \label{eq:gamam_cond_5}
\end{eqnarray}
which comes from the lowest order term of the Magnus expansion of the time-evolution operator~\cite{Schutjens13, Theis16}; the derivation is given in Appendix~\ref{sec:a_wahwah}.
Equation~\ref{eq:gamam_cond_3} ensures that, given the $|11\rangle$ qubit state, the coupler state is $|e\rangle$ at $t=t_g/2$, and Eqs.~\ref{eq:gamam_cond_4} and~\ref{eq:gamam_cond_5} ensure that, given the other logical states, the coupler state returns to $|g\rangle$ at $t=t_g/2$. This approximation introduces an error, accounting for gate infidelity of $\sim 10^{-3}$ for the optimized $\text{CZ}$ gate found using this approach. However, this is not a dominant contribution to the total gate infidelity found in Sec.~\ref{sec:4}.

We choose the test pulse for the first-half of the gate, $\Omega_0(t)$, to be:
\begin{eqnarray}
    \Omega_0(t) &= & \mathcal{A}e^{-\frac{\left(t-t_g/4\right)^2}{2\sigma^2}}\left[1-\cos\left(\omega_{1}\left(t-\frac{t_g}{4}\right)\right)\right] \nonumber \\ && \times\left[1-\cos\left(\omega_{2}\left(t-\frac{t_g}{4}\right)\right)\right],  \\ && \quad \text{(for $0\leq  t \leq \frac{t_g}{2}$)} \nonumber
\end{eqnarray}
which we refer to as Gaussian-shaped Multitone Amplitude Modulation (GaMAM).
The pulse is similar to the WAHWAH (Weak Anharmonicity with Average Hamiltonian) pulse suggested in earlier works \cite{Schutjens13, Theis16}. One difference is that we use two modulation frequencies $\omega_1$ and $\omega_2$. The other difference is that $\Omega_y(t)$ is not defined by the time-derivative of $\Omega_x(t)$, unlike Derivative Removal by Adiabatic Gate (DRAG) schemes, including WAHWAH~\cite{Motzoi09}.

Using numerical optimization techniques, the specific pulse parameters $\mathcal{A}$, $\sigma$, $\omega_1$, and $\omega_2$ that satisfy Eqs.~\ref{eq:gamam_cond_3},~\ref{eq:gamam_cond_4}, and ~\ref{eq:gamam_cond_5} are found. The appropriate $\Theta$ that satisfies Eq.~\ref{eq:gamam_cond_5} is found after the choice of $\Omega_0(t)$. Through the optimization process, we present the $\sqrt{\mathrm{CZ}}$ gate with gate time $t_g = 16/\Delta \omega_c$ as shown in Figs.~\ref{fig:sqrt_cz_pulse_envelope} and~\ref{fig:sqrt_cz_pulse_envelope_freq}. It turns out, $\omega_1\sim\omega_2\sim 3\Delta \omega_c$ minimizes coherent errors, as it produces zeros at the coupler frequency conditioned by $|01\rangle$, $|10\rangle$, and $|11\rangle$ states (Fig.~\ref{fig:sqrt_cz_pulse_envelope_freq}).
The resulting evolution of the coupler, conditioned by different qubit states, is shown in Figs.~\ref{fig:sqrt_cz_sig_z_trajectory} and~\ref{fig:sqrt_cz_bloch}. The pulse parameters found are presented in Appendix~\ref{sec:a_table}.

\subsection{\label{sec:3.3} Mitigating Effects of Quasiparticle Tunneling}
When performing gates with superconducting circuits, a quasiparticle might tunnel across the Josephson junction, altering $n^C_g$ by 0.5~\cite{Ristè13}. As our coupler is sensitive to the external charge, this would lead to the change of both $\omega_c$ and $\Delta \omega_c$. Even if this process occurs on a timescale much larger than a single gate operation, it could hamper the measurement process as a whole.

For our gate, a dual-tone drive may be considered, where we simultaneously apply two drive pulses, with each $\omega_d$ and $\Omega_0(t)$ defined from different choices of $\omega_c$ and $\Delta \omega_c$. One drive-pulse component would target the coupler with the DC gate charge of $n^C_{g,0}$, while the other would target $n^C_{g,0}+ 0.5$, which are the gate charges with and without quasiparticle tunneling. Since each pulse targets an individual gate charge, the difference between $\omega_c(n^C_g = n^C_{g,0})$ and $\omega_c(n^C_g = n^C_{g,0}+ 0.5)$ should be larger than all other frequency scales. At the bias point shown in Fig.~\ref{fig:frequency_shift}, $\omega_c/2\pi$ differs by at least $\SI{1}{GHz}$ for the two quasiparticle parities, thereby satisfying this requirement.

\subsection{\label{sec:3.4} Hyperfine Magnetic Noise and Microwave Photon Loss}

Thus far in this manuscript, we have primarily focused on dephasing from charge noise, due to the inherent susceptibility of the system: faster gates with higher $\Delta \omega_c$ require more charge sensitivity. While other decoherence mechanisms may dominate in different systems or during single-qubit operations, they are slow compared to the gate times $t_g \lesssim \SI{50}{ns}$ obtained in Sec.~\ref{sec:4}. Hyperfine magnetic noise is associated with $T^*_2 \gtrsim \SI{1}{\micro s}$ for spin qubits in isotopically pure Si/SiGe devices~\cite{Kawakami14, Kerckhoff21, Bo24}.
Superconducting resonators grown directly on semiconductor substrates in hybrid configurations are generally plagued by microwave photon loss, with quality factors on the order of $10^3$~\cite{Mi17}, corresponding to
$T_1$ of order $\SI{1}{\micro s}$ for typical GHz-scale resonator frequencies.
Moreover, flip-chip architectures where the quantum dots and the coupler are on separate tiers could lead to much higher Q factors on the order of $10^4$~\cite{Holman21, Corrigan23}.
While these lifetimes are small compared to those in realizations of high-Q superconducting resonators~\cite{Mahashabde20}, they are nevertheless considerably longer than $t_g$.
Furthermore, our coupler is only transiently populated during the gate; it is not fully populated during most of the gate operation for the case of pulse-envelope-engineered CZ gates. Thus, we estimate $< 10^{-4}$ infidelity from hyperfine noise and $< 10^{-2}$ infidelity from photon loss.

\section{\label{sec:4} Expected CZ Gate Fidelity under $1/f^\beta$ Charge Noise}

In this section, we estimate the CZ gate fidelity in the presence of $1/f^\beta$ charge noise for various charge-noise powers and differential lever arms $\alpha^D$. We provide estimates with and without dynamical decoupling, as dynamical decoupling exposes the coupler to charge noise for longer periods of time, even if it is known to reduce the qubit dephasing rate approximately by a factor of 4, for the echo sequence~\cite{Ithier05}. The simulation results show that dynamical decoupling may be required when the low-frequency charge noise on spin qubits is relatively stronger than the charge noise on the coupler. 
Unless otherwise noted, we set $\beta=1$ for the analytics and simulations.

Charge noise acts on both the qubits and the coupler, complicating fidelity calculations. As long as the effect of decoherence is small enough, we approximate the total gate infidelity from decoherence to be a linear sum of the two gate infidelity contributions. We determine the infidelity from qubit dephasing using the standard analytic model for undriven systems~\cite{Ithier05}. We calculate infidelity from coupler dephasing through a time-domain numerical simulation based on a cumulant expansion.
Time-domain simulation is needed, because the conventional constant decay rate model cannot capture the dynamics of a driven system for the first few Rabi oscillations; recent work suggests that the conventional model overestimates the decoherence~\cite{Groszkowski23}. Our approach does not require averaging over random noise samples, making it computationally efficient.
Note that dephasing during the adiabatic transition between the charge-insensitive DFS regime and the charge-sensitive RX regime is not included in our simulations.
Finally, we note that although the pulse shape was derived using the lowest-order Magnus expansion, our simulations do not make this approximation and therefore capture any coherent infidelity associated with it. 
As noted above, we find this infidelity to be approximately $\sim 10^{-3}$ for a dynamically decoupled gate and zero for an off-resonant gate.

Details of our analysis of $1/f$ charge noise dephasing are provided in Appendices~\ref{sec:a_dephasing} and~\ref{sec:a_F}. We show that the infidelity from qubit dephasing is, to first order, independent of the quantum-dot parameters.
Similarly, the infidelity from coupler dephasing is only weakly dependent on the coupler parameters.
Therefore, we optimize quantum-dot parameters to minimize coupler dephasing, and we optimize coupler parameters to minimize qubit dephasing.

\begin{figure}[ht!]
\subfloat{\label{fig:delta_n_g}}
\subfloat{\label{fig:slope}}
\includegraphics{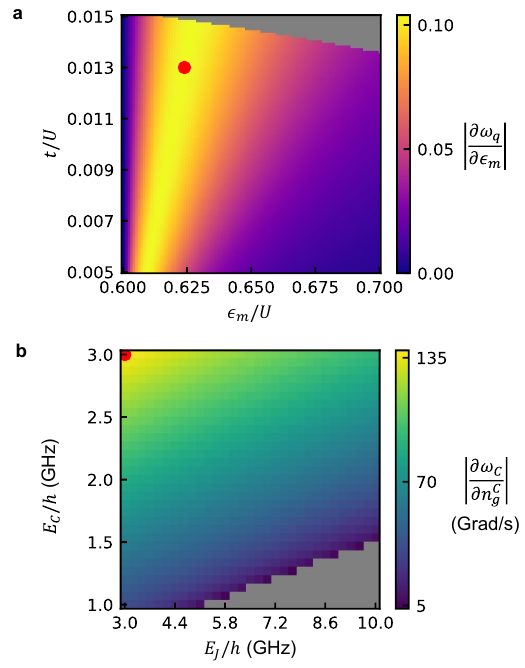}
\caption{\textbf{Expected values of the charge sensitivity of the qubit ($\frac{\partial \omega_q}{\partial \epsilon_m}$) and coupler ($\frac{\partial \omega_c}{\partial n^C_g}$}). \textbf{(a)}~Expected values of $\frac{\partial \omega_q}{\partial \epsilon_m}$, calculated from the analytic model of RX qubit. The grey area is the area where the exchange energy $J/h > \SI{0.7}{GHz}$. The maximum is noted as a red point, with $\frac{\partial \omega_q}{\partial \epsilon_m} = 0.104$. \textbf{(b)}~Expected values of $\frac{\partial \omega_c}{\partial n^C_g}$, calculated from the charge basis representation of the OCS transmon. For each values of $E_J$ and $E_C$, the DC bias $n^C_{g,0}$ is set such that $\omega_c(n^C_g = n^C_{g,0})/2\pi=\omega_c(n^C_g = n^C_{g,0}+ 0.5)/2\pi \pm 1$ GHz, to take into account of quasiparticle tunneling. 
The grey area is where the circuit is too insensitive to the offset charge and therefore has no $n^C_{g,0}$ that satisfies the condition.
Between $\frac{\partial \omega_c}{\partial n^C_g}(n^C_g=n^C_{g,0})$ and $\frac{\partial \omega_c}{\partial n^C_g}(n^C_g=n^C_{g,0}+0.5)$, the larger value is chosen.
The maximum is noted as a red point, with $\frac{\partial \omega_c}{\partial n^C_g} = 139 \text{~Grad/s}$. Note that this exact value differs from the rough estimate $4E_C/\hbar = 151 \text{~Grad/s}$ by 8\%.} 
\label{fig:fig5}  
\end{figure}

The analysis in Appendix~\ref{sec:a_dephasing} shows that the maximum gate fidelity is achieved by maximizing the differential lever arm $\alpha^D$ and the charge sensitivity of the qubit $\partial \omega_q/\partial \epsilon_m$ and of the coupler $\partial \omega_c/\partial n^C_g$. We restrict the parameter space to experimentally plausible values. Capacitance between the OCS transmon coupler and ground usually constrains $E_C$; we set $E_C/h \leq 3$ GHz. A typical Josephson junction fabrication process constrains $E_J/h \geq 2\sim3$ GHz~\cite{Lu23}. 
On the spin qubit side, we constrain $\epsilon_m$ and $t$ so that the exchange energy is $J/h=2t^2U/(U^2-\epsilon_m^2)\leq\SI{0.7}{GHz}$; for large values of $J$, the electron wavefunctions have a large overlap along the quantum dots and the Fermi-Hubbard model is invalid~\cite{Pan20}. The Fermi-Hubbard parameters for Si-SiGe quantum dots are set to be $U=4$ meV and $U_C=0.2U$. The values of $\partial \omega_q/\partial \epsilon_m$ and $\partial \omega_c/\partial n^C_g$ within this parameter space is shown in Fig.~\ref{fig:fig5}. We proceed with fidelity estimations taking $\partial \omega_q/\partial \epsilon_m = 0.104$ and $\partial \omega_c/\partial n^C_g = 139~$Grad/s. Note that the specific coupler parameter choice for the targeted coupler charge sensitivity is $E_J/h = E_C/h = 3$ GHz. This lies near the regime of the Cooper-pair box, characterized by $E_J/E_C \sim 0.1$~\cite{Nakamura99, Devoret04}. However, we emphasize that this parameter choice is based on the assumption of strict $1/f$ noise with $\beta = 1$. The optimal parameter depends on the specific noise profile, and our model remains valid across a broader regime with increased $E_J/E_C$ and reduced charge sensitivity, as illustrated in Fig.~\ref{fig:slope}.

The infidelity from charge noise strongly depends on the charge-noise power.
We use a double-sided noise spectrum convention, for which earlier works suggest reference noise power spectral density of $A^D_0 = \SI{0.21}{\micro eV^2/Hz}$~\cite{Elliot22} on the qubit detuning $\epsilon_m$ and $A^C_0 = \SI{0.5}{(10^{-3}e)^2/Hz}$~\cite{Krantz19} on coupler gate charge $q^C_g = (2e)n^C_g$.
Also, we take the low-frequency cutoff for the $1/f$ noise as $\omega_l/2\pi = \SI{10}{kHz}$.

\begin{figure*}[ht!]
\includegraphics{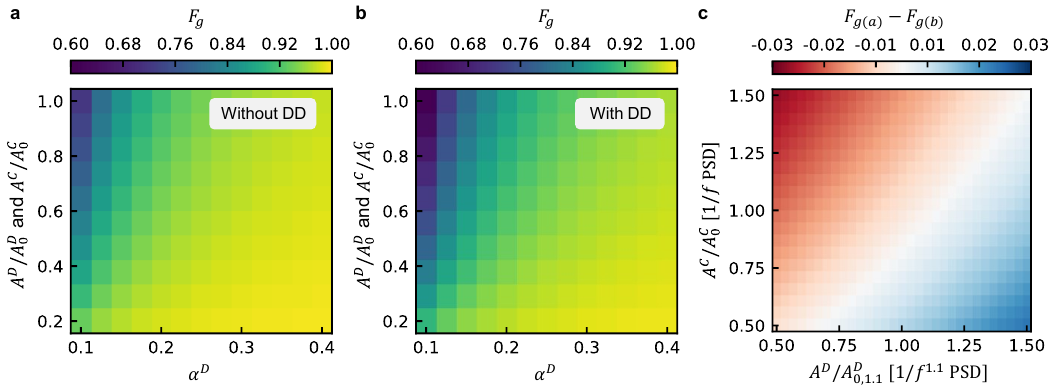}
\subfloat{\label{fig:cz_no_dd_F}}
\subfloat{\label{fig:cz_dd_F}}
\subfloat{\label{fig:cz_F_beta}}
\caption{\textbf{Comparing gate fidelity without and with dynamical decoupling.} \textbf{(a)}~CZ gate fidelity without dynamical decoupling, in the presence of $1/f$ noise. The noise power ratio and the lever arm are varied. The noise power relative to the base noise powers $A^D_0 = \SI{0.21}{\micro eV^2/Hz}$ and $A^C_0 =  \SI{0.5}{(10^{-3}e)^2/Hz}$. CZ gate fidelity of 91\% is expected for the base noise power and the base differential lever arm $\alpha^D=0.2$. \textbf{(b)}~CZ gate fidelity with dynamical decoupling, in the presence of $1/f$ noise. The noise power ratio and the lever arm are varied, with the same noise power ratio as Fig.~\ref{fig:cz_no_dd_F}. \textbf{(c)}~Difference in fidelity without and with dynamical decoupling (blue color indicates the gate without dynamical decoupling performs better). Fidelities are calculated under $1/f^{1.1}$ charge noise on quantum dots, $1/f$ coupler charge noise, and $\alpha^D=0.2$.
The base qubit charge noise power for $1/f^{1.1}$ noise profile, $A^D_{0, 1.1}$, is adjusted from $A^D_0$ as fit parameters of the charge noise spectroscopy depend on the exponent $\beta$. $A^D_{0, 1.1} = A^D_0[10~\text{MHz}/1~\text{Hz}]^{0.1}$ is chosen so that the two different representations lead to the same noise power at 10 MHz, which is a typical crossover point based on the comparison of different reports of $1/f^\beta$ charge noise sources~\cite{Astafiev04, Gustafsson13}. 
As low-frequency charge noise on the quantum dots becomes the prevailing noise source, dynamical decoupling becomes advantageous.}
\label{fig:fig6}  
\end{figure*}

In Fig.~\ref{fig:cz_no_dd_F} and Fig.~\ref{fig:cz_dd_F}, we present the expected gate fidelity of the CZ gates without and with dynamical decoupling, respectively. 
Fidelities are presented as a function of $\alpha^D$ and charge noise power; here, the charge noise of the qubits and the coupler are varied together under a $1/f$ noise profile. 
Under the assumption of $1/f$ charge noise, the fidelity is higher for the gate without dynamical decoupling, regardless of the charge noise power.
For our best guess of typical parameters $A^D=A^D_0$, $A^C=A^C_0$, and $\alpha^D=0.2$ (which is within the range of values from earlier reports~\cite{Beaudoin16, Harvey18, Borjans20_splitgate, Zwerver22}), we predict 91\% gate fidelity. The corresponding $\Delta \omega_c/2\pi$ and $g^D/2\pi$ are $\SI{230}{MHz}$ and $\SI{52}{MHz}$, respectively, with the gate time of $t_g\lesssim \SI{10}{ns}$ for both gate types.

Interestingly, the higher-performing gate sequence depends on the exact profile of the dot and coupler charge noise.
In a realistic device, the noise profile on the spin qubits and the coupler may differ, particularly in flip-chip architectures. 
In Fig.~\ref{fig:cz_F_beta}, we present the difference in fidelity between the gate without dynamical decoupling and the gate with dynamical decoupling under the assumption of $1/f^\beta$ charge noise with $\beta=1$ for the coupler and $\beta=1.1$~\cite{Zwerver22} for the quantum dots.
The choice $\beta=1.1$ means that, compared to a $1/f$ profile, the noise power is shifted to low-frequencies~\cite{Zwerver22, Gustafsson13}.
Indeed, when low-frequency dot noise is more significant, the dynamical decoupling sequence is advantageous.
When coupler noise is more significant, or when there is more high-frequency dot noise (i.e. $\beta\leq 1$), the faster gate time provided by the gate without dynamical decoupling outperforms.
Thus, the adoption of dynamical decoupling should depend on materials properties.

\section{\label{sec:5} Conclusion}
In this work, we propose a CZ MAP gate mediated by an OCS transmon coupler to generate entanglement between two distant quantum-dot spin qubits. 
The MAP gate is enabled by a shift in the coupler frequency depending on the logical state of the spin qubits.
Two different gate sequences are investigated, a faster protocol without dynamical decoupling and a slower one with dynamical decoupling. To accommodate a pulse sequence for a dynamically decoupled CZ gate, we provide a systematic approach to engineer the pulse envelope for CPhase($\theta$) gate with arbitrary conditional phase $\theta$.

Charge noise is the main source of incoherent gate error. 
To accurately determine the dephasing of RX-regime qubits and a driven coupler under $1/f^\beta$ noise profiles, we presented time-domain simulations based on a cumulant expansion. Analytics reveal that the qubit dephasing is independent of quantum dot parameters to first order, and, for $\beta=1$, coupler dephasing is insensitive to coupler parameters.
Using realistic estimates for device parameters, gate fidelities above 90\% are expected,
which is on the scale of the error correction threshold for connective elements between processor modules~\cite{Ramette24}.
Furthermore, increasing the differential lever arm and decreasing the charge noise power would enable fidelities exceeding 99\%.
When the charge noise on the spin qubits is more severe or more weighted towards low frequencies than the charge noise on the coupler, the dynamically decoupled gate outperforms the faster off-resonant gate.

Several hardware advancements could further improve the gate performance: notably, reducing the coupler-to-ground capacitance would enable higher $E_C$.
Additionally, as we expect charge noise to be the dominant source of error, improving the quality of the electrostatic gates for the quantum dots would significantly enhance gate performance.
With these improvements, OCS transmon couplers and nonlinear superconducting circuits in general could become viable quantum interconnects between quantum dot spin qubit modules for a scalable quantum computing architecture.

\begin{acknowledgments}
The authors are grateful to Joe Kerckhoff, Nathan Holman, Andrew Pan, Marc Dvorak, and Andrew Hunter for helpful discussions on quantum dots-superconducting circuits hybrid systems.
The authors are also grateful to Peter Groszkowski and Antoine Brilliant for valuable suggestions on time-domain simulations of non-Markovian noise.
HHK is supported by Korea Foundation for Advanced Studies.
ITR and MH are supported by an appointment to the Intelligence Community Postdoctoral Research Fellowship Program at the Massachusetts Institute of Technology administered by Oak Ridge Institute for Science and Education (ORISE) through an interagency agreement between the U.S. Department of Energy and the Office of the Director of National Intelligence (ODNI).
\end{acknowledgments}

\appendix
\section{\label{sec:a_table} Table of the Key Parameters}
A summary of the key parameters used for numerical analysis is given in Table~\ref{tab:table1}.

\begin{widetext}
\begin{table*}
\caption{\label{tab:table1}\textbf{Table of the key parameters.} The parameters used in time-domain simulations are given.}
\renewcommand{\arraystretch}{1.2}
\begin{ruledtabular}
\begin{tabular}{lrll}
\textrm{Symbol, Quantity}&
\textrm{Value}&
\textrm{Units}&
\textrm{Notes}\\
\colrule
Fixed quantum dot parameters, & &  \\
$U^D$, On-site energy & 4 & \SI{}{meV} & \\
$U_c^D$, Off-site energy & 0.8 & \SI{}{meV} & $=0.2U^D$\\
$\alpha^D$, Base differential lever arm & 0.2 & & \\
\\
Optimized qubit parameters, & & \\
$\epsilon_m^D$, Dimple detuning energy &  2.496 & \SI{}{meV}& $=0.624 U^D$\\
$\epsilon^D$, Left-right detuning energy &  0&  meV & $=0$ for RX qubits\\
$t_{ij}^D$, Tunneling energy & 0.014 &  meV& $=0.013 U^D$\\
$\omega_q$, Qubit frequency & 7.61 & $(2\pi)$\SI{}{GHz} & \\
$\left|\partial \omega_q/\partial \epsilon_m\right|$, Qubit charge sensitivity & 0.104 & & \\
\\
Optimized coupler parameters, & & \\
$E_J$, Josephson energy & 3 & $h\cdot$\SI{}{GHz} & \\
$E_C$, Charging energy & 3 & $h\cdot$\SI{}{GHz} & \\
$n^C_{g,0}$, Coupler gate charge DC bias & 0.326 & & \makecell[l]{$n^C_{g,0}+0.5$ also used due to \\ quasiparticle tunneling}\\
$\omega_{c,0}$, Coupler frequency without coupling & 5.16 & $(2\pi)$\SI{}{GHz} & using $\alpha^D_1=\alpha^D_2=\alpha^D_3=0$\\
$\omega_{c,\ket{11}}$, Coupler frequency for qubit state $\ket{11}$& 5.81& $(2\pi)$\SI{}{GHz} & using $\alpha^D=0.2$\\
$\left|\partial \omega_c/\partial n^C_g\right|$, Coupler charge sensitivity & 139 & \SI{}{Grad/s} &\\
\\
Coupling parameters, & & \\
$\Delta n^C_g$, Coupler gate charge shift & 0.0104 & & $=(\alpha^D/2) |\partial \omega_q/\partial \epsilon_m|$ \\
$ \Delta \omega_c$, Coupler frequency shift & 230 & $(2\pi)$\SI{}{MHz} & $=\Delta n^C_g |\partial \omega_c/\partial n^C_g|$\\
\\
Off-resonant CZ gate parameters, \\
$\delta$, Drive detuning &$-115$ & $(2\pi)$\SI{}{MHz} & $=-\Delta \omega_c/2$ \\
$\Omega_x$, In-phase drive amplitude & 148 &$(2\pi)$\SI{}{MHz}& $=\sqrt{5/12} \Delta \omega_c$ \\
$\Omega_y$, Out-phase drive amplitude & 0 &$(2\pi)$\SI{}{MHz}& \\
$t_g$, Gate time & 5.33 & \SI{}{ns} & $=\pi\sqrt{6}/\Delta\omega_c$ \\
\\
Pulse-envelope-engineered $\sqrt{\text{CZ}}$ gate parameters, & & \\
$\delta$, Drive detuning & 0 & $(2\pi)$\SI{}{MHz} & resonant driving \\
$A$, GaMAM amplitude & 82.3 & $(2\pi)$\SI{}{MHz} & $=0.358\Delta \omega_c$\\
$\sigma$, GaMAM standard deviation & 1.87 & \SI{}{ns} & $=2.7/\Delta \omega_c$\\
$\omega_1$, First GaMAM modulation frequency & 643 & $(2\pi)$\SI{}{MHz} & $=2.8\Delta \omega_c$\\
$\omega_1$, Second GaMAM modulation frequency & 712 & $(2\pi)$\SI{}{MHz} & $=3.1\Delta \omega_c$\\
$\Theta$, Azimuthal rotation angle for the second half of the drive & 0.402 & \SI{}{rad} & \\
$t_g$, Gate time & 11.1 & \SI{}{ns} & $=16/\Delta\omega_c$ \\
\\
Noise parameters, & & \\
$A^D_0$, Base quantum dot noise power & 0.21 & \SI{}{\micro eV^2/Hz} &\\
$A^C_0$, Base coupler noise power & 0.5 & \SI{}{(10^{-3}e)^2/Hz} & \\
\end{tabular}
\end{ruledtabular}
\end{table*}
\end{widetext}

\section{\label{sec:a_rx} RX Qubit Analytics}

In this section, we provide analytic expressions on RX qubits, including the expressions for the eigenstates, the qubit frequency $\omega_q$, the charge sensitivity $\partial \omega_q/\partial \epsilon_m$, and the longitudinal coupling strength $g^D$.

\subsection{Fermi-Hubbard Model \\ of Triple Quantum Dot System}
\begin{figure}[ht!]
\includegraphics{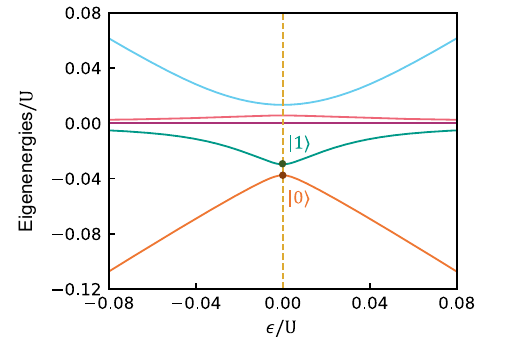}
\caption{\textbf{Energy levels of a triple quantum dot system.} 
The eigenenergies of a triple quantum dot system with $U_C = 0.2U$, $\epsilon_m = 0.624U$, and $t^D_{ij} = t = 0.13U$ are plotted as a function of the detuning parameter $\epsilon$. A large magnetic field is applied to exclude states with negative $S_z$ from the plot. The eigenenergies are offset by the energy of the $\ket{Q_{1/2}}$ state, with quantum numbers $S = 3/2$ and $S_Z = 1/2$, represented by the magenta line. The $\ket{Q_{1/2}}$ state does not couple to the computational subspace in ideal noiseless conditions. The lowest two energy levels in the plot form the qubit logical basis ($\{\ket{0}^D, \ket{1}^D\}$). The bias point for RX qubits is $\epsilon = 0$, denoted by the dotted yellow line.}
\label{fig:figA2}  
\end{figure}

The Fermi-Hubbard model description of a triple quantum dot system is (Eq.~\ref{eq:fh}):
\begin{eqnarray}
    \hat{H}_{\mathrm{FH}} &= &\frac{U^D}{2} \sum_{i} \hat{n}^D_{i} (\hat{n}^D_{i}-1) + U_c^D \sum_{\langle i,j \rangle} \hat{n}^D_{i} \hat{n}^D_{j} + \sum_{i} V^D_i\hat{n}^D_{i} \nonumber \\
    &&- \sum_{\langle i,j \rangle, \hat{\sigma}}t_{ij}^D \left( \hat{c}^{\dagger D}_{i\sigma} \hat{c}^D_{j\sigma} + \text{H.c.} \right).
\end{eqnarray}
It is useful to define the detuning parameters~\cite{Russ17}, shown in Fig.~\ref{fig:rx_qubit}:
\begin{eqnarray}
    \epsilon &= &\frac{\mu_1 - \mu_3}{2} = \frac{V^D_1 - V^D_3}{2} = 0, \\
    \epsilon_m& = &\mu_2 - \frac{(\mu_1 + \mu_3)}{2} = V^D_2 - \frac{(V^D_1 + V^D_3)}{2} + U_C,
\end{eqnarray}
where $\mu_i$ is the chemical potential of the $i$th dot. EO qubits operate at $\epsilon = 0$ in the DFS encoding and in the RX regime. Note that some other works define detuning parameters via gate voltages rather than chemical potentials~\cite{Pan20}.
By diagonalizing Eq.~\ref{eq:fh}, the energy levels of the triple quantum dot system can be numerically calculated, as shown in Fig.~\ref{fig:figA2}.

The EO logical subspace is formed by three-electron states with $S=S_z=1/2$. Permitting double site occupancy, there are eight such number states; specifically, the logical states in the RX regime are primarily superpositions of the four such low energy states:
\begin{eqnarray}
    |T\rangle^D&=&\sqrt{\frac{2}{3}}|t_+\rangle_{13}|\!\downarrow\rangle_2 - \frac{1}{\sqrt{3}}|t_0\rangle_{13}|\!\uparrow\rangle_2 \nonumber\\
    &=&\frac{1}{\sqrt{6}}\left(2|\!\uparrow,\downarrow,\uparrow\rangle-|\!\uparrow,\uparrow,\downarrow\rangle-|\!\downarrow,\uparrow,\uparrow\rangle\right),\\
    |S\rangle^D&=&|s\rangle_{13}|\!\uparrow\rangle_2=\frac{1}{\sqrt{2}}\left(|\!\uparrow,\uparrow,\downarrow\rangle-|\!\downarrow,\uparrow,\uparrow\rangle\right),\\
    |L\rangle^D&=&|s\rangle_{11}|\!\uparrow\rangle_3=|\!\uparrow\downarrow,0,\uparrow\rangle,\\
    |R\rangle^D&=&|\!\uparrow\rangle_1|s\rangle_{33}=|\!\uparrow,0,\uparrow\downarrow\rangle,
\end{eqnarray}
due to the large chemical potential of the middle quantum dot. $|s\rangle_{ii}=|\!\!\uparrow\downarrow\rangle_{i}$ is the two-electron state occupying a single dot, single orbital. The Pauli exclusion principle forces the state to be a singlet state. $|s\rangle_{ij}=(|\!\uparrow\rangle_i|\!\downarrow\rangle_j-|\!\downarrow\rangle_i|\!\uparrow\rangle_j)/\sqrt{2}$ is the two-electron singlet state occupying two quantum dots $i$ and $j$. $|t_0\rangle_{ij}=(|\!\uparrow\rangle_i|\!\downarrow\rangle_j+|\!\downarrow\rangle_i|\!\uparrow\rangle_j)/\sqrt{2}$ and $|t_+\rangle_{ij}=|\!\uparrow\rangle_i|\!\uparrow\rangle_j$ are the two-electron triplet states for the two quantum dots.

Evaluating the matrix elements of the Fermi-Hubbard Hamiltonian Eq.~(2.2) gives the Hamiltonian of the subspace:
\begin{equation}
    \hat{H}_{\mathrm{FH}} = \begin{pmatrix}
    0 & 0 & -\frac{\sqrt{6}}{2}t^D_{12} & \frac{\sqrt{6}}{2}t^D_{23} \\
    0 & 0 & -\frac{\sqrt{2}}{2}t^D_{12} & -\frac{\sqrt{2}}{2}t^D_{23} \\
    -\frac{\sqrt{6}}{2}t^D_{12} & -\frac{\sqrt{2}}{2}t^D_{12} & \Delta_{FH} & 0 \\
    \frac{\sqrt{6}}{2}t^D_{23} & -\frac{\sqrt{2}}{2}t^D_{23} & 0 & \Delta_{FH}
    \end{pmatrix}, \label{eq:fh_matrix}
\end{equation}
in the basis $\{|T\rangle, |S\rangle, |L\rangle, |R\rangle\}^D$. Coupling between $|T\rangle^D$ and $|S\rangle^D$ is mediated by a two-step electron tunneling via $|L\rangle$ and $|R\rangle^D$. The $|L\rangle^D$ and $|R\rangle^D$ states have the same energies $\Delta_{FH}=U^D-2U^D_C-\epsilon_m$, relative to the energies of $|T\rangle^D$ and $|S\rangle^D$ states, as $\epsilon=0$.

\subsection{Eigenstates and Eigenenergies of RX Qubit}
Diagonalizing Eq.~\ref{eq:fh_matrix} gives the eigenstates and eigenenergies. The qubit states of the RX qubit are the two lowest-energy states:
\begin{widetext}
\begin{eqnarray}
    |0\rangle^D&=&\frac{1}{\sqrt{4\Delta_{FH}^2+48t^2-4\Delta_{FH}\sqrt{\Delta_{FH}^2+12t^2}}}\left[2\sqrt{6}t|T\rangle^D+(\sqrt{\Delta_{FH}^2+12t^2}-\Delta_{FH})(|L\rangle^D-|R\rangle^D)\right], \label{eq:fh_eigenstate0} \\
    |1\rangle^D&=&\frac{1}{\sqrt{4\Delta_{FH}^2+16t^2-4\Delta_{FH}\sqrt{\Delta_{FH}^2+4t^2}}}\left[2\sqrt{2}t|S\rangle^D+(\sqrt{\Delta_{FH}^2+4t^2}-\Delta_{FH})(|L\rangle^D+|R\rangle^D)\right],\label{eq:fh_eigenstate1}
\end{eqnarray}
\end{widetext}
where we set the tunneling energy between the adjacent dots to be equal (i.e., $t^D_{12}=t^D_{23}=t$).

The eigenenergies for the four states $\omega_{|i\rangle}$ and the qubit energy $\omega_q$ are:
\begin{eqnarray}
    \omega_{|0\rangle}&=&\frac{\Delta_{FH}-\sqrt{\Delta_{FH}^2 + 12t^2}}{2}, \\ 
    \omega_{|1\rangle}&=&\frac{\Delta_{FH}-\sqrt{\Delta_{FH}^2 + 4t^2}}{2},  \\
    \omega_{|2\rangle}&=&\frac{\Delta_{FH}+\sqrt{\Delta_{FH}^2 + 4t^2}}{2},\\
    \omega_{|3\rangle}&=&\frac{\Delta_{FH}+\sqrt{\Delta_{FH}^2 + 12t^2}}{2},  \\
    \omega_q&=&\omega_{|1\rangle}-\omega_{|0\rangle} \nonumber \\
    &=& \frac{1}{2}\left(\sqrt{\Delta_{FH}^2 + 12t^2}-\sqrt{\Delta_{FH}^2 + 4t^2}\right) .\label{eq:FH_energies}
\end{eqnarray}
\subsection{Analytic Expressions \\for RX Qubit Sensitivity $\partial \omega_q/\partial \epsilon_m$  }

The qubit sensitivity $\partial \omega_q/\partial \epsilon_m$ is a key parameter of our model. Taking a partial derivative of Eq.~\ref{eq:FH_energies} gives:
\begin{eqnarray}
    \frac{\partial \omega_q}{\partial \epsilon_m}=\frac{1}{2}\left(\frac{\Delta_{FH}}{\sqrt{\Delta_{FH}^2+4t^2}}-\frac{\Delta_{FH}}{\sqrt{\Delta_{FH}^2+12t^2}}\right). \label{eq:fh_qubit_sensitivity_1}
\end{eqnarray}

\begin{figure}[t!]
\includegraphics{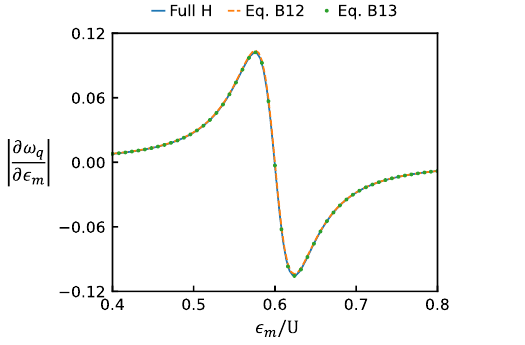}
\caption{\textbf{Comparison of different equations for qubit charge sensitivity, $\frac{\partial \omega_q}{\partial \epsilon_m}$, for $t = 0.013U$.} The three methods of calculating $\frac{\partial \omega_q}{\partial \epsilon_m}$ are compared: calculating the difference of eigenvalues of the full Hamiltonian ($\hat{H}_{\text{FH}}$), Eq.~\ref{eq:fh_qubit_sensitivity_1}, and Eq.~\ref{eq:fh_qubit_sensitivity_2}. The three show a good match.}
\label{fig:figA1}  
\end{figure}

Alternatively, the qubit energy may be written in terms of the expectation values of the charge operator $\hat{n}^D_2$. An intuitive explanation is that, in the RX regime, $\epsilon_m$ is much bigger than other energy scales, therefore, the eigenstates are only changed marginally for the perturbation of $\epsilon_m$. Consequently, the qubit sensitivity is proportional to the difference of the number of electrons in the middle dot. We could explicitly show this through the Hellmann-Feynman theorem:
\begin{widetext}
\begin{eqnarray}
    \frac{\partial \omega_q}{\partial \epsilon_m} &= & \frac{\partial}{\partial \epsilon_m}\left(\langle 1|\hat{H}_{\mathrm{FH}}|1\rangle^D-\langle 0|\hat{H}_{\mathrm{FH}}|0\rangle^D\right) \nonumber\\
    &= & \left( \langle1|\frac{\partial \hat{H}_{\mathrm{FH}}}{\partial \epsilon_m}|1\rangle^D-\langle0|\frac{\partial \hat{H}_{\mathrm{FH}}}{\partial \epsilon_m}|0\rangle^D\right)+\left[\left(\frac{\partial}{\partial \epsilon_m}\langle 1|\right)\hat{H}_{\mathrm{FH}}|1\rangle^D-\left(\frac{\partial}{\partial \epsilon_m}\langle 0|\right)\hat{H}_{\mathrm{FH}}|0\rangle^D\right] \nonumber\\
    &&+ \left[\langle 1|\hat{H}_{\mathrm{FH}}\left(\frac{\partial}{\partial \epsilon_m}|1\rangle^D\right)-\langle 0|\hat{H}_{\mathrm{FH}}\left(\frac{\partial}{\partial \epsilon_m}|0\rangle^D\right)\right]\nonumber\\
    &= & \left( \langle1|\frac{\partial \hat{H}_{\mathrm{FH}}}{\partial \epsilon_m}|1\rangle^D-\langle0|\frac{\partial \hat{H}_{\mathrm{FH}}}{\partial \epsilon_m}|0\rangle^D\right)+\left[\omega_1\left(\frac{\partial}{\partial \epsilon_m}\langle 1|\right)|1\rangle^D-\omega_0\left(\frac{\partial}{\partial \epsilon_m}\langle 0|\right)|0\rangle^D\right] \nonumber\\
    &&+\left[\omega_1\langle 1|\left(\frac{\partial}{\partial \epsilon_m}|1\rangle^D\right)-\omega_0\langle 0|\left(\frac{\partial}{\partial \epsilon_m}|0\rangle^D\right)\right] \nonumber\\
    &= & \langle1|\frac{\partial \hat{H}_{\mathrm{FH}}}{\partial \epsilon_m}|1\rangle^D-\langle0|\frac{\partial \hat{H}_{\mathrm{FH}}}{\partial \epsilon_m}|0\rangle^D = \langle 1|\hat{n}^D_2|1\rangle^D-\langle 0|\hat{n}^D_2|0\rangle^D,\label{eq:fh_qubit_sensitivity_2}
\end{eqnarray}
\end{widetext}
where we have used:
\begin{eqnarray}
\left(\frac{\partial}{\partial \epsilon_m}\bra{i}\right)\ket{i}^D+\bra{i}\left(\frac{\partial}{\partial \epsilon_m}\ket{i}^D\right)  = \frac{\partial}{\partial \epsilon_m} \langle i | i \rangle^D = 0,
\end{eqnarray}
for $i=0,1$.
By plugging in the expression for logical qubit states Eq.~\ref{eq:fh_eigenstate0} and Eq.~\ref{eq:fh_eigenstate1}, we recover the same result as Eq.~\ref{eq:fh_qubit_sensitivity_1}.
Eq.~\ref{eq:fh_qubit_sensitivity_2} is used to derive the analytic expression for the infidelities from qubit dephasing, in Appendix~\ref{sec:a_dot_dephasing}. The different methods to calculate the qubit sensitivity show a good match in the RX regime, as verified in Fig.~\ref{fig:figA1}.

Finally, using the above two equations, the longitudinal coupling strength $g^D$ may be expressed in terms of Fermi-Hubbard parameters:
\begin{eqnarray}
    g^D &\simeq & 2E_C n^C_\text{ZPF}\alpha^D \left(\langle0|\hat{n}^D_2|0\rangle^D-\langle1|\hat{n}^D_2|1\rangle^D\right) \nonumber\\
    &= & E_C n^C_\text{ZPF}\alpha^D \nonumber \\&&\times \left(\frac{\Delta_{\text{FH}} }{\sqrt{\Delta_{FH}^2+12t^2}}-\frac{\Delta_{\text{FH}} }{\sqrt{\Delta_{FH}^2+4t^2}}\right)^D.
\end{eqnarray}

\begin{figure*}[ht!]
\includegraphics{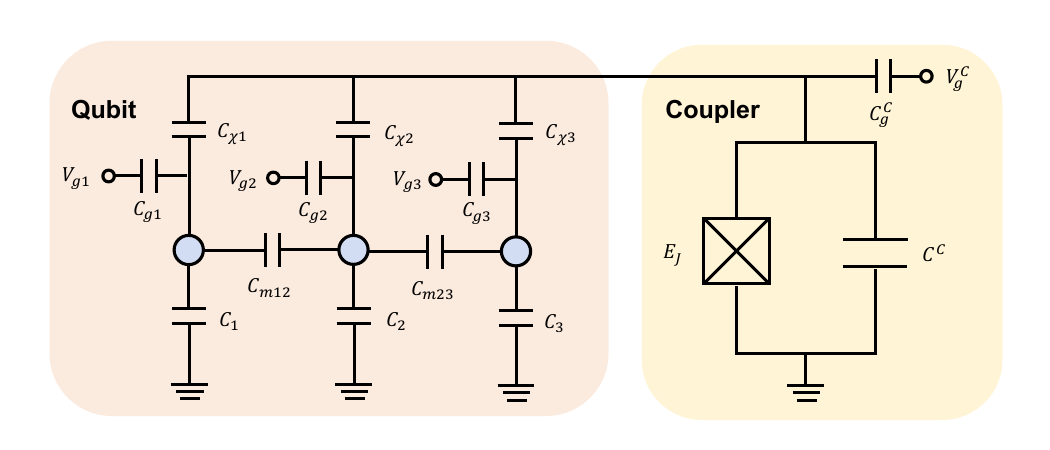}
\caption{\textbf{Circuit diagram of the triple dot spin qubit - OCS transmon hybrid system}. A lumped element circuit model of the hybrid system. Each quantum dots are coupled to the ground and the coupler with capacitance $C_i$ and $C_{\chi i}$, for $i=1,2,3$. The neighboring quantum dots $i$ and $j$ are coupled with each other with $C_{mij}$. Each quantum dots are coupled to the control voltage source $V_{gi}$, with coupling capacitance $C_{gi}$. The OCS transmon coupler is coupled to the ground with capacitance $C^C$ and the inductive energy of the Josephson junction is $E_J$. The coupler is coupled to the control voltage source $V_g^C$ with coupling capacitance $C^C_g$. Note that the diagram assumes an additional coupling gate between the qubit and the coupler~\cite{Borjans20_splitgate, Bottcher22}, which would be absent in an architecture employing direct coupling through the plunger gate line~\cite{Landig18, Landig19, Collard22, Dijkema25}. In the latter case, the grounded OCS transmon shown in the schematic should be replaced with a differential OCS transmon to enable a sufficient DC bias on the quantum dots.
} \label{fig:hybrid_system_schematic}
\end{figure*}

\section{\label{sec:a_quantization} Quantization of the \\ Triple Dot-OCS Transmon Hybrid System}

In this section, the interaction Hamiltonian in Sec.~\ref{sec:2.1} is derived. For simplicity, we consider a single triple dot qubit coupled to an OCS transmon coupler. Assuming minimal direct interaction between the two spin qubits, the interaction Hamiltonian is the sum of the interaction between the coupler and each qubit. The circuit diagram of the system is shown in Fig.~\ref{fig:hybrid_system_schematic}. 

We follow the canonical quantization, starting from classical variables. The charge of each dots $Q_i$ for dots $i=1,2,3$ and the coupler charge $Q^C$ are expressed as:
\begin{eqnarray}
    Q_1 &= & C_1 V_1 + C_{g1}(V_1 - V_{g1}) + C_{\chi 1}(V_1 - V^C) \nonumber \\ &&+ C_{m12}(V_1 - V_2), \\
    Q_2 &= & C_2 V_2 + C_{g2}(V_2 - V_{g2}) + C_{\chi 2}(V_2 - V^C)\nonumber \\ && + C_{m12}(V_2 - V_1) + C_{m23}(V_2 - V_3), \\ 
    Q_3 &= & C_3 V_3 + C_{g3}(V_3 - V_{g3}) + C_{\chi 3}(V_3 - V^C)\nonumber \\ && + C_{m23}(V_3 - V_2), \\
    Q^C &= & C^C V^C + C_{\chi 1}(V^C - V_1) + C_{\chi 2}(V^C - V_2) \nonumber \\ &&  + C_{\chi 3}(V^C - V_3) + C_g^C(V^C - V_g^C).
\end{eqnarray}

We express the node charge and voltage variables in vector form, offset by the control gate voltages:
\begin{eqnarray}
    \textbf{Q}:=\begin{bmatrix}
        Q_1 + C_{g1}V_{g1} \\
        Q_2 + C_{g2}V_{g2} \\
        Q_3 + C_{g3}V_{g3} \\
        Q^C + C_g^C V_g^C
    \end{bmatrix},
    \textbf{V}:=\begin{bmatrix}
        V_1 \\
        V_2 \\
        V_3 \\
        V^C
    \end{bmatrix}.
\end{eqnarray}
This allows the capacitive network to be expressed as:
\begin{eqnarray}
    \textbf{Q} = \textbf{C}\textbf{V},
\end{eqnarray}
where $\textbf{C}$ is the capacitance matrix of the system:
\begin{equation}
    \textbf{C} = \begin{pmatrix}
        C_{\Sigma 1} & -C_{m12} & 0 & C_{\chi1} \\
        -C_{m12} & C_{\Sigma 2} & -C_{m23} & C_{\chi2} \\
        0 & -C_{m23} & C_{\Sigma 3} & C_{\chi3} \\
        C_{\chi1} & C_{\chi2} & C_{\chi3} & C^C_\Sigma
    \end{pmatrix}.
\end{equation}
Here, $C_{\Sigma i} = C_i + \sum_j C_{mij} + C_{\chi i} + C_{gi}$ is the total capacitance of the $i$th dot, and $C^C_\Sigma= C^C + \sum_i C_{\chi i} + C^C_g$ represents the total capacitance of the coupler node.

The energy stored in capacitive elements is:
\begin{eqnarray}
    \mathcal{T}_E = \frac{1}{2} \textbf{V}^T \textbf{C}\textbf{V} = \frac{1}{2} \textbf{V}^T \textbf{Q} = \frac{1}{2} \textbf{Q}^T \textbf{C}^{-1}\textbf{Q}. 
\end{eqnarray}

The inductive energy of the system $\mathcal{T}_M$ is the energy across the Josephson Junction:
\begin{eqnarray}
    \mathcal{T}_M = - E_J \cos(\phi^C),
\end{eqnarray}
where $E_J$ is the inductive energy of the Josephson junction, and $\phi^C$ is the reduced flux, or the phase across the junction:
\begin{eqnarray}
    \phi^C = \frac{2\pi\Phi^C}{\Phi_0} = \frac{2\pi}{\Phi_0}\int_{-\infty}^t{V^C(t') dt},
\end{eqnarray}
where $\Phi^C$ is the flux across the junction and $\Phi_0 = h/(2e)$ is the superconducting flux quantum. 

The Legendre transformation of the Lagrangian $\mathcal{L}=\mathcal{T}_E - \mathcal{T}_M$ yields the classical Hamiltonian of the system $H_{\text{EM}}$ without interdot tunneling between quantum dots:
\begin{eqnarray}
    H_{\text{EM}} = \frac{1}{2}\textbf{Q}^T \textbf{C}^{-1}\textbf{Q} - E_J \cos(\phi^C).
\end{eqnarray}

The classical interaction Hamiltonian $H_\textrm{int}$ is the energy term dependent on both coupler and quantum dot variables. It is determined by the $(i,\, 4)$ element of $\textbf{C}^{-1}$:
\begin{eqnarray}
    H_{\textrm{int}} = \sum_{i=1,2,3} \left[\textbf{C}^{-1}\right]_{i4}Q_iQ^C.
\end{eqnarray}

Canonical quantization yields the quantum mechanical interaction Hamiltonian:
\begin{eqnarray}
    \hat{H}_{\textrm{int}} = \sum_{i=1,2,3} \left[\textbf{C}^{-1}\right]_{i4}\hat{Q}_i\hat{Q}^C. \label{eq:int_H_appendix}
\end{eqnarray}

Equation~\ref{eq:int_H_appendix} respects the full capacitance network, which renormalizes the charge variables and thus gives an accurate value of the interaction strength. The Hamiltonian is simplified under the realistic assumption that the capacitance to ground is dominant for each nodes: $C_i, C^C \gg C_{\chi i}, C_{mij}$. The leading terms of the $(i,\, 4)$ elements of the inverse of the capacitance matrix are:
\begin{eqnarray}
    \left[\textbf{C}^{-1}\right]_{14} \simeq \frac{1}{C_1 C_2 C_3 C^C}C_2 C_3 C_{\chi1} = \frac{C_{\chi1}}{C_1 C^C}, \\
    \left[\textbf{C}^{-1}\right]_{24} \simeq \frac{1}{C_1 C_2 C_3 C^C}C_1 C_3 C_{\chi2} = \frac{C_{\chi2}}{C_2 C^C},\\
    \left[\textbf{C}^{-1}\right]_{34} \simeq \frac{1}{C_1 C_2 C_3 C^C}C_1 C_2 C_{\chi3} = \frac{C_{\chi3}}{C_3 C^C}.
\end{eqnarray}

The interaction Hamiltonian is then approximated as:
\begin{eqnarray}
    \hat{H}_{\textrm{int}} &\simeq & \sum_{i=1,2,3} 2e^2 \frac{C_{\chi i}}{C_i C^C} \hat{n}_i \hat{n}^C \\
    &= & \sum_{i=1,2,3} 4 E_C \alpha^D_i \hat{n}_i \hat{n}^C,
\end{eqnarray}
where $\hat{n}_i = \hat{Q}_i/e$, $\hat{n}^C = \hat{Q}^C/(2e)$, and $\alpha^D_i = C_{\chi i}/C_i$ is the lever arm of the $i$th dot. By adding the superscript $D$ for quantum dot variables and summing over the two qubits $D=A,B$, we get the interaction Hamiltonian of Eq.~\ref{eq:intH_2}.

\section{\label{sec:a_longitudinal} Longitudinal Coupling Condition \\ for RX Qubits}
In this section, the longitudinal coupling condition for RX qubits:
\begin{equation}
\alpha^D_1 = \alpha^D_3,
\end{equation}
is derived. Then, it will be shown that Eq.~\ref{eq:intH_2} is expressed as Eq.~\ref{eq:intH_3} with Eq.~\ref{eq:g_d} under the longitudinal coupling condition.

First, we re-express Eq.~\ref{eq:intH_2} in the form:
\begin{eqnarray}
\hat{H}_\text{int} &=& \sum_{D=A,B} 4E_C\hat{n}^C \underbrace{\left(\sum_{i=1,2,3} \alpha^D_i \hat{n}^D_i \right)}_{(*)}.
\end{eqnarray}

We are interested in expressing $(*)$ in the qubit Pauli operators for qubits $D=A,B$:
\begin{eqnarray}
\hat{\sigma}^A_k &=& \hat{\sigma}_k \otimes \hat{I} \otimes \hat{I}, \\
\hat{\sigma}^B_k &=& \hat{I} \otimes \hat{\sigma}_k \otimes \hat{I},
\end{eqnarray}
acting on the state $\ket{\psi} = \ket{\psi}_A \otimes \ket{\psi}_B \otimes \ket{\psi}_C$. We want to calculate each element:
\begin{eqnarray}
\langle 0 |(*)| 0 \rangle^D &=& \sum_{i=1,2,3} \alpha^D_i \langle 0 |\hat{n}_i | 0 \rangle^D  \\
\langle 0 |(*)| 1 \rangle^D &=& \sum_{i=1,2,3} \alpha^D_i \langle 0 |\hat{n}_i | 1 \rangle ^D \label{eq:transition_matrix_dot_charge}\\
\langle 1 |(*)| 1 \rangle^D &=& \sum_{i=1,2,3} \alpha^D_i \langle 1 |\hat{n}_i | 1 \rangle^D .
\end{eqnarray}

To evaluate the matrix elements, we consider the spin-charge basis $\{\ket{T}, \ket{S}, \ket{L}, \ket{R}\}^D$ for the three-electron subspace used in Appendix~\ref{sec:a_rx}, which are also the eigenstates of the dot electron number operator. The states $\ket{T}^D$ and $\ket{S}^D$ have one electron per dot, yielding:
\begin{eqnarray}
\hat{n}_1\ket{T}^D &=& \hat{n}_2\ket{T}^D = \hat{n}_3\ket{T}^D = \ket{T}^D, \label{eq:n_in_tslr_1}\\
\hat{n}_1\ket{S}^D &=& \hat{n}_2\ket{S}^D = \hat{n}_3\ket{S}^D = \ket{S}^D. \label{eq:n_in_tslr_2}
\end{eqnarray}
The states $\ket{L}^D$ and $\ket{R}^D$ have charge configurations of $(2,0,1)$ and $(1,0,2)$, respectively:
\begin{eqnarray}
\hat{n}_1\ket{L}^D &=& 2\ket{L}^D \nonumber\\
\hat{n}_2\ket{L}^D &=& 0, \nonumber\\
\hat{n}_3\ket{L}^D &=& \ket{L}^D, \label{eq:n_in_tslr_3}
\end{eqnarray}
and
\begin{eqnarray}
\hat{n}_1\ket{R}^D &=& \ket{R}^D, \nonumber\\
\hat{n}_2\ket{R}^D &=& 0, \nonumber\\
\hat{n}_3\ket{R}^D &=& 2\ket{R}^D.\label{eq:n_in_tslr_4}
\end{eqnarray}
Thus, each $\hat{n}_i\ket{0}^D$ and $\hat{n}_i\ket{1}^D$ may be found using Eqs.~\ref{eq:n_in_tslr_1},~\ref{eq:n_in_tslr_2},~\ref{eq:n_in_tslr_3},~\ref{eq:n_in_tslr_4},~\ref{eq:fh_eigenstate0}, and~\ref{eq:fh_eigenstate1}. This leads to:
\begin{eqnarray}
\langle 0 |\hat{n}_1 | 1\rangle^D   &=& - \langle 0 |\hat{n}_3 | 1\rangle^D  , \\
\langle 0 | \hat{n}_2 | 1 \rangle^D   &=& 0.
\end{eqnarray}
For a purely longitudinal interaction, Eq.~\ref{eq:transition_matrix_dot_charge} should be zero. Thus, $\alpha^D_1 = \alpha^D_3$ leads to a purely longitudinal interaction.

Next, we derive the interaction Hamiltonian in the qubit Pauli basis (Eq.~\ref{eq:intH_3}) from that in the charge basis (Eq.~\ref{eq:intH_2}), with the coupling strength given as Eq.~\ref{eq:g_d}. 
Using that the total number of electrons in the qubit sums up to 3 and the condition $\alpha^D_1 = \alpha^D_3$, we have:
\begin{eqnarray}
\langle 0 |(*)| 0 \rangle^D
&=& 2\alpha_1^D \left(\frac{3- \langle 0 |\hat{n}_2 | 0 \rangle^D}{2}\right)  + \alpha_2^D\langle 0 |\hat{n}_2 | 0 \rangle^D \nonumber \\
&=& 3 \alpha^D_1 + (\alpha^D_2 - \alpha^D_1)\langle 0 |\hat{n}_2 | 0 \rangle^D, 
\end{eqnarray}
and similarly,
\begin{eqnarray}
\langle 1 |(*)| 1 \rangle^D
&=& 3 \alpha^D_1 + (\alpha^D_2 - \alpha^D_1)\langle 1 |\hat{n}_2 | 1 \rangle^D.
\end{eqnarray}
Thus, the interaction Hamiltonian acting on the qubit subspace is:
\begin{widetext}
\begin{eqnarray}
\hat{H}_{\text{int}} &=& \sum_{D=A,B} 4E_C \hat{n}^C \left(\sum_{i=1,2,3} \alpha^D_i \hat{n}^D_i \right) \nonumber \\
&=& 4 E_C \hat{n}^C \sum_{D=A,B} \left[\sum_{v, w = 0, 1} \Bigl(\langle v | (*) | w \rangle \Bigr)\ket{v}\!\bra{w}\right]^D \nonumber \\
&=& 4 E_C \hat{n}^C \sum_{D=A,B} \left[ \left(3 \alpha^D_1 + (\alpha^D_2 - \alpha^D_1)\langle 0 |\hat{n}_2 | 0 \rangle^D\right) \frac{\hat{I} +\hat{\sigma}^D_z}{2}
+ \left( 3 \alpha^D_1 + (\alpha^D_2 - \alpha^D_1)\langle 1 |\hat{n}_2 | 1 \rangle^D\right) \frac{\hat{I} -\hat{\sigma}^D_z}{2} \right] \nonumber \\
&=& 4E_C \hat{n}^C \sum_{D=A,B}\left[ \frac{ \alpha^D\left(\langle 1 |\hat{n}_2 | 1 \rangle^D + \langle 1 |\hat{n}_2 | 1 \rangle^D\right) + 6\alpha^D_1}{2}\hat{I}
+ \frac{ \alpha^D\left(\langle 1 |\hat{n}_2 | 1 \rangle^D - \langle 1 |\hat{n}_2 | 1 \rangle^D\right)}{2}\hat{\sigma}^D_z
\right] \\
&=& \sum_{D=A,B} 2E_C n^C_{\text{ZPF}}\left[\left\{\alpha^D\left(\langle 0 |\hat{n}_2 | 0 \rangle^D + \langle 1 |\hat{n}_2 | 1 \rangle^D\right) + 6\alpha^D_1\right\} \hat{I} + \left\{\alpha^D\left(\langle 0 |\hat{n}_2 | 0 \rangle^D - \langle 1 |\hat{n}_2 | 1 \rangle^D\right)\right\} \hat{\sigma}^D_z \right]\frac{\hat{n}^C}{n^C_{\text{ZPF}}},
\end{eqnarray}
\end{widetext}
where we have introduced the differential lever arm $\alpha^D = \alpha^D_1 - \alpha^D_2 = \alpha^D_3 - \alpha^D_3$. Thus, we recover the longitudinal interaction Hamiltonian of Eq.~\ref{eq:intH_3}, with the coupling strengths $g^D$ and $g^D_I$ given as Eqs.~\ref{eq:g_d} and~\ref{eq:g_I}.

\section{\label{sec:a_drive_matrix_representation} Driven System in the Rotating Frame}
In this section, we explicitly present the Hamiltonian of the driven system in the rotating frame, $\hat{H}_R(t)$, which is used for the time-domain noise simulation of the system in Appendix~\ref{sec:a_dephasing}.
We apply the basis transform $\hat{U}_\mathrm{RW}(t)$:
\begin{eqnarray}
    \hat{U}_\mathrm{RW}(t) &=& \prod_{a,b=0,1} e^{i\omega_d t |abe\rangle\langle abe|} \nonumber \\
    &=& \sum_{a,b=0,1} \ket{abg}\bra{abg} + e^{i\omega_d t} \ket{abe}\bra{abe},
\end{eqnarray}
to the sum of transverse drive Hamiltonian $\hat{H}_\mathrm{drive}(t)$ given in Eq.~\ref{eq:drive_H_lab} and the lab frame bare Hamiltonian $\hat{H}_0
= \sum_{a,b=0,1}\sum_{c=g,e} \omega_{abc}\ket{abc}\bra{abc}$, to yield the rotating frame Hamiltonian $\hat{H}_R$:
\begin{eqnarray}
\hat{H}_R(t) &=& \hat{U}_\mathrm{RW} \left(\hat{H}_0 + \hat{H}_{\mathrm{drive}}(t)\right) \hat{U}^{\dagger}_\mathrm{RW} - i\hat{U}_\mathrm{RW}\frac{d}{dt}\hat{U}^{\dagger}_\mathrm{RW} \nonumber \\
&=& \sum_{a,b=0,1} \omega_{abg} \ket{abg}\bra{abg} + (\omega_{abe}-\omega_d)\ket{abe}\bra{abe} \nonumber \\
&& \quad + \frac{\Omega_x(t)}{2} \hat{\sigma}^C_x + \frac{\Omega_y(t)}{2} \hat{\sigma}^C_y,
\end{eqnarray}
where
\begin{eqnarray}
    \hat{\sigma}^C_x &=& \sum_{a,b=0,1}|abe\rangle\langle abg| + |abg\rangle\langle abe|, \\
    \hat{\sigma}^C_y &=& i\sum_{a,b=0,1}|abe\rangle\langle abg| - |abg\rangle\langle abe|,
\end{eqnarray}
are the coupler Pauli operators.

In matrix notation, $\hat{H}_R$ is represented as:
\begin{widetext}
\begin{equation}
    \hat{H}_{R}(t) =  
    \begin{bmatrix}
    \omega_{00g} & \Omega(t)/2 & 0 & 0 & 0 & 0 & 0 & 0 \\
    \Omega^*(t)/2 & \omega_{00e} - \omega_d & 0 & 0 & 0 & 0 & 0 & 0 \\
    0 & 0 & \omega_{01g} & \Omega(t)/2 & 0 & 0 & 0 & 0 \\
    0 & 0 & \Omega^*(t)/2 & \omega_{01e} - \omega_d & 0 & 0 & 0 & 0 \\
    0 & 0 & 0 & 0 & \omega_{01e} & \Omega(t)/2 & 0 & 0 \\
    0 & 0 & 0 & 0 & \Omega^*(t)/2 & \omega_{01g} - \omega_d & 0 & 0 \\
    0 & 0 & 0 & 0 & 0 & 0 & \omega_{11g} & \Omega(t)/2 \\
    0 & 0 & 0 & 0 & 0 & 0 & \Omega^*(t)/2 & \omega_{11e} - \omega_d \\
    \end{bmatrix}, \label{eq:matrix_H_raw}
\end{equation}
\end{widetext}
where $\Omega(t) = \Omega_x(t) - i\Omega_y(t)$ is the complex drive amplitude. We work in the basis corresponding to the Kronecker product of the Pauli basis of each logical space $|\psi\rangle = |\psi\rangle_A \otimes |\psi\rangle_B \otimes |\psi\rangle_C$, such that the order is $|00g\rangle,\, |00e\rangle,\, |01g\rangle,\, \text{etc.}$.
As the logical states of the qubits are unaltered, $\hat{H}_R(t)$ is block-diagonal in the four subspaces corresponding to the two-qubit logical states.

For our CPhase($\theta$) gates, Eq.~\ref{eq:matrix_H_raw} simplifies further. 
For our coupler gate bias $n^C_{g,0}$, $\omega_{11g}-\omega_{10g}\simeq \omega_{01g}-\omega_{00g}$. Thus, the phase acquisition from bare qubit frequencies $\omega_{abg}$ approximately cancel when calculating $\theta$ in Eq.~\ref{eq:central_phase_eq}, so that we may neglect the frequency offsets $\omega_{abg}$. This is also reflected in Fig.~\ref{fig:drive_principle}.
Note that the phase acquisition from bare qubit frequencies can also be accounted for by adjusting the drive angle $\Theta$ in the pulse-envelope-engineering scheme.
Furthermore, the transition frequencies differ by $\Delta \omega_c$ when the system is operated at the linear regime. These points together lead to the simplified expression for $\hat{H}_R(t)$:
\begin{widetext}
\begin{equation}
    \hat{H}_{R}(t) \rightarrow  
    \begin{bmatrix}
    0 & \Omega(t)/2 & 0 & 0 & 0 & 0 & 0 & 0 \\
    \Omega^*(t)/2 & \delta + 2\Delta \omega_c & 0 & 0 & 0 & 0 & 0 & 0 \\
    0 & 0 & 0 & \Omega(t)/2 & 0 & 0 & 0 & 0 \\
    0 & 0 & \Omega^*(t)/2 & \delta + \Delta \omega_c & 0 & 0 & 0 & 0 \\
    0 & 0 & 0 & 0 & 0 & \Omega(t)/2 & 0 & 0 \\
    0 & 0 & 0 & 0 & \Omega^*(t)/2 & \delta + \Delta \omega_c & 0 & 0 \\
    0 & 0 & 0 & 0 & 0 & 0 & 0 & \Omega(t)/2 \\
    0 & 0 & 0 & 0 & 0 & 0 & \Omega^*(t)/2 & \delta \\
    \end{bmatrix}\label{eq:driven_H_RWA},
\end{equation}
\end{widetext}
where $\delta=\omega_{c,|11\rangle} - \omega_d$ is the detuning between the drive and the transition frequency, of the $|11\rangle$ subspace. Without loss of generality, we assume that the $\omega_{c,|11\rangle}$ subspace has the lowest frequency, as shown in the inset of Fig.~\ref{fig:frequency_shift}. Otherwise, the sign in front of $\Delta \omega_c$ will trivially change.

\section{\label{sec:a_DD} Dynamical Decoupling Pulse Derivation}

Here, we show that the pulse sequence presented in Fig.~\ref{fig:DD_sequence} leads to a CZ gate operation, up to single qubit rotations. For simplicity, we trace out the coupler state and leave only the two qubits, as ideally the coupler returns to the ground state after each step. Then the unitary evolution from $\sqrt{\mathrm{CZ}}$ gate in the matrix notation, $\hat{U}_{\sqrt{\mathrm{CZ}}}$, is:

\begin{equation}
    \hat{U}_{\sqrt{\mathrm{CZ}}}=\begin{bmatrix}
1&0&0&0\\
0&1&0&0\\
0&0&1&0\\
0&0&0&e^{i\pi/2}
\end{bmatrix}_{AB}.
\end{equation}
The single qubit X gate in matrix notation is:
\begin{equation}
    \hat{\sigma}_{xx}=\begin{bmatrix}
0&1\\
1&0
\end{bmatrix}_A\otimes
\begin{bmatrix}
0&1\\
1&0
\end{bmatrix}_B
=\begin{bmatrix}
0&0&0&1\\
0&0&1&0\\
0&1&0&0\\
1&0&0&0
\end{bmatrix}_{AB}.
\end{equation}
The total evolution is:
\begin{eqnarray}
    \hat{\sigma}_{xx}\hat{U}_{\sqrt{\mathrm{CZ}}}\hat{\sigma}_{xx}\hat{U}_{\sqrt{\mathrm{CZ}}} & = &
\begin{bmatrix}
e^{i\pi/2}&0&0&0\\
0&1&0&0\\
0&0&1&0\\
0&0&0&e^{i\pi/2}

\end{bmatrix}_{AB},
\end{eqnarray}
which satisfies Eq.~\ref{eq:central_phase_eq} for a CZ gate: $\theta=(\pi/2-0)-(0-\pi/2)=\pi$.

\section{Magnus Expansion of the Gate Unitary \label{sec:a_wahwah}}
The conditions Eqs.~\ref{eq:gamam_cond_3},~\ref{eq:gamam_cond_4}, and~\ref{eq:gamam_cond_5} are derived by transforming the rotating wave Hamiltonian $\hat{H}_R(t)$ of Eq.~\ref{eq:driven_H_RWA} into the interaction frame, where the basis transformation $\hat{U}_I(t)$ is chosen to make the diagonal elements of the Hamiltonian zero, similar to the approach shown in Refs.~\cite{Schutjens13, Theis16}. This is explicitly given as:
\begin{eqnarray}
\hat{U}_I(t) &=& \Bigl[\ket{11g}\bra{11g} + \exp\bigl(i \delta t\bigr)\ket{11e}\bra{11e}\Bigr] \nonumber \\
&& + \Bigl[\ket{10g}\bra{10g} + \exp\bigl(i(\delta + \Delta \omega_c)t\bigr)\ket{10e}\bra{10e}\Bigr] \nonumber \\
&& + \Bigl[\ket{01g}\bra{01g} + \exp\bigl(i (\delta + \Delta \omega_c)t\bigr)\ket{01e}\bra{01e}\Bigr] \nonumber \\
&& + \Bigl[\ket{00g}\bra{00g} + \exp\bigl(i(\delta + 2\Delta \omega_c)t\bigr)\ket{00e}\bra{00e}\Bigr], \nonumber \\
\end{eqnarray}
where we started with the general case in which the carrier frequency of the drive could be detuned from the $\ket{11g} \leftrightarrow \ket{11e}$ transition (i.e., $\delta \neq 0$).

The interaction Hamiltonian in the resulting frame is:
\begin{eqnarray}
\hat{H}_I(t) &=& \hat{U}_I \hat{H}_R(t) \hat{U}_I^\dagger - i\hat{U}_I\frac{d\hat{U}_I^\dagger}{dt} \nonumber \\
&=& \frac{\Omega(t)}{2}\Bigl[e^{-i\delta t}\ket{11g}\bra{11e} + e^{-i(\delta + \Delta \omega_c) t}\ket{10g}\bra{10e} \nonumber \\
&& + e^{-i(\delta + \Delta \omega_c) t}\ket{01g}\bra{01e} +
e^{-i(\delta + 2\Delta \omega_c) t}\ket{00g}\bra{00e} \Bigr] \nonumber \\
&& + \text{H.c.}. \label{eq:interaction_H_WAHWAH}
\end{eqnarray}
The time-evolution operator for the system at $t = t_g/2$ is:
\begin{eqnarray}
\hat{U}\left(t = \frac{t_g}{2}\right) = \mathcal{T}\text{exp}\left(-i\int^\frac{t_g}{2}_0 dt \hat{H}_I(t) \right), \label{eq:time_order}
\end{eqnarray}
where $\mathcal{T}$ is the time-ordering operator, required for the general case where $\hat{H}_I(t)$ does not commute at different times (i.e., $\left[\hat{H}_I(t_1), \hat{H}_I(t_2)\right] \neq 0$). However, this generally makes a closed analytic form for $\hat{U}$ intractable. The time-ordering operator can be removed by expanding the time evolution into an infinite series~\cite{Theis16}:
\begin{eqnarray}
\hat{U}\left(t = \frac{t_g}{2}\right) = \text{exp}\left(-i \sum_k\hat{\mathcal{M}}_k\left(t = \frac{t_g}{2}\right)\right), \label{eq:magnus_expansion}
\end{eqnarray}
where $\hat{\mathcal{M}}_k$ are terms in the Magnus expansion. The first two terms are:
\begin{widetext}
\begin{eqnarray}
\hat{\mathcal{M}}_0\left(t = \frac{t_g}{2}\right) &=& \int^\frac{t_g}{2}_0 dt \hat{H}_I(t), \label{eq:magnus_zeroth} \\
\hat{\mathcal{M}}_1\left(t = \frac{t_g}{2}\right) &=& -\frac{i}{2}\int^\frac{t_g}{2}_0 dt_2 \int^\frac{t_g}{2}_0 dt_1 \left[\hat{H}_I(t_2), \hat{H}_I(t_1)\right]. \label{eq:magnus_first}
\end{eqnarray}
\end{widetext}

In this work, only the lowest order term $\hat{\mathcal{M}}_0$ is considered when implementing a gate, as the approximation error from neglecting higher-order terms is not a limiting factor of the gate fidelity. Explicitly substituting Eq.~\ref{eq:interaction_H_WAHWAH} and Eq.~\ref{eq:magnus_zeroth} into Eq.~\ref{eq:magnus_expansion} gives:
\begin{widetext}
\begin{eqnarray}
\hat{U}\left(t=\frac{t_g}{2}\right) \simeq & ~\text{exp} \Bigg( & -\underbrace{i\int^\frac{t_g}{2}_0 dt \frac{\Omega(t)}{2}e^{-i\delta t}\ket{11g}\bra{11e}}_{\text{evolution between $\ket{11g}$ and $\ket{11e}$}}  -\underbrace{i\int^\frac{t_g}{2}_0 dt \frac{\Omega(t)}{2}e^{-i(\delta + \Delta \omega_c)t} \ket{10g}\bra{10e}}_{\text{evolution between $\ket{10g}$ and $\ket{10e}$}} \nonumber \\
&& -\underbrace{i\int^\frac{t_g}{2}_0 dt \frac{\Omega(t)}{2}e^{-i(\delta + \Delta \omega_c)t} \ket{01g}\bra{01e}}_{\text{evolution between $\ket{01g}$ and $\ket{01e}$}} -\underbrace{i\int^\frac{t_g}{2}_0 dt \frac{\Omega(t)}{2}e^{-i(\delta + 2\Delta \omega_c)t} \ket{00g}\bra{00e}}_{\text{evolution between $\ket{00g}$ and $\ket{00e}$}} + \text{H.c.} \Bigg).
\end{eqnarray}
\end{widetext}
The condition given as Eq.~\ref{eq:gamam_cond_1} requires full population inversion between $\ket{11g}$ and $\ket{11e}$ at $t = t_g/2$, while returning to the initial state (i.e., the coupler ground state) otherwise. This means that the evaluation of the integrals inside the underbraces should yield $\pi$ for the first underbrace and 0 for the others. By setting $\delta = 0$ and $\Omega(t) = \Omega_0(t)$, the constraints given in Eqs.~\ref{eq:gamam_cond_3}, \ref{eq:gamam_cond_4}, and \ref{eq:gamam_cond_5} are recovered. It is worth highlighting that these three constraints only determine the population of the ground and excited coupler states at $t = t_g/2$. The remaining constraints, provided in Eqs.~\ref{eq:omega_x}, \ref{eq:omega_y}, and \ref{eq:gamam_cond_2}, complete the $\sqrt{\text{CZ}}$ gate with the gate time $t_g$ and the acquired phase $\theta = \pi/2$.

\section{Dephasing from $1/f$ Charge Noise \label{sec:a_dephasing}}
In this section, the governing equations for the dephasing from $1/f$ charge noise are given. As mentioned in the main text, realistic charge noise follows a $1/f^\beta$ noise profile with a range of $\beta$ over different frequency regimes. For simplicity, in this section we assume strictly $1/f$ noise with a low-frequency cutoff. The results show that the optimal circuit and quantum dot parameters to minimize gate infidelities are those that increase the charge sensitivity.

\subsection{\label{sec:4.1} RX Qubit Dephasing\label{sec:a_dot_dephasing}}
The decoherence of a non-driven qubit in the lab frame, when the noise is coupled to the quantization axis, is well-studied. \cite{Ithier05} It is expected to follow a Guassian decay, and we can directly use the formula for the dephasing rate $\Gamma_2^D$.

The noise Hamiltonian $\hat{H}^D_n$ in the presence of a longitudinal $1/f$ noise is expressed as:
\begin{eqnarray}
    \hat{H}^D_n & = & \zeta^A(t) \sum_{b=0,1}\sum_{c=g,e} |1bc\rangle\langle1bc| \nonumber \\
     && + \zeta^B(t) \sum_{a=0,1}\sum_{c=g,e} |a1c\rangle\langle a1c|,
\end{eqnarray}
where $\zeta^D(t)$ is the classical noise on the frequency for qubit $D=A,B$. The first term corresponds to the charge noise on qubit A, and the second term explains the charge noise on qubit B. Gaussian $1/f$ qubit frequency noise is described by the double-sided noise spectral density $S^D[\omega]$, which is the Fourier transform of the autocorrelation of $\zeta^D(t)$:
\begin{eqnarray}
    S^D[\omega] = \begin{cases}
        A^D_\omega \frac{2\pi}{|\omega|} &\text{if $\omega_l < |\omega| <\omega_h$} \\
        0 &\text{otherwise}
    \end{cases},
\end{eqnarray}
where $A^D_\omega$ is the power spectral density at $\SI{1}{Hz}$. $\omega_l$ and $\omega_h$ are the low- and high-frequency cutoffs, respectively. We assume that the noise on the two qubits are uncorrelated but have the same power: $A^A_\omega = A^B_\omega$.

For sufficiently large $\omega_h$, the decay rates in the first order are:
\begin{eqnarray}
    \Gamma_2^D  &=&
    \begin{cases}
        \Gamma_{2,0}^D(t) &\text{(without DD)} \\
        \Gamma_{2,\text{echo}}^D &\text{(with DD)}
    \end{cases}, \\
    \Gamma^D_{2,0}(t)&= &\sqrt{A^D_\omega }\sqrt{\text{ln}\left(\frac{1}{\omega_l t}\right)+O(1)}, \\
    \Gamma^D_{2,\text{DD}}&= & \sqrt{A^D_\omega }\sqrt{\text{ln}(2)}.
\end{eqnarray}
$\Gamma^D_{2,0}(t)$ is the decay rate without dynamical decoupling. It is a weak function of time, only dependent on the square root of the logarithm. $\Gamma^D_{2,\text{DD}}$ is the decay rate with dynamical decoupling. The factor of $\sqrt{\text{ln}(2)}$ changes slightly for $1/f^\beta$ noise with $\beta \neq 1$, but not by a substantial amount as long as $\beta \sim 1$~\cite{Harvey18}.

The charge noise on frequency in the first order is related both to the charge sensitivity of the qubit $\partial \omega_q/\partial \epsilon_m$ and the charge noise on the qubit detuning:
\begin{eqnarray}
    A^D_\omega = \left(\frac{\partial \omega_q^D}{\partial \epsilon_m}\right)^2 A^D
\end{eqnarray}
where $\omega^D_q$ is the qubit frequency and $A^D$ is the charge noise power spectral density on $\epsilon_m$, at 1 Hz, as mentioned in the main text. Substituting the base noise power for the spin qubits, $A^D_0$, yields $T^D_{2,\text{echo}} = \SI{104}{ns}$ for the optimal qubit parameters chosen in the main text. While $T^{*D}_2 = 1/\Gamma^D_{2,0}$ is a weak function of time, its time-averaged value over $[0, t_g = \pi\sqrt{6}/\Delta \omega_c]$ for the off-resonant CZ gate, with a cutoff frequency of $\omega_l = \SI{10}{kHz}$ and differential lever arm $\alpha^D = 0.2$, is approximately $\SI{29}{ns}$.

Note that perturbation at the symmetrical detuning $\epsilon$ does not affect the qubit frequency in first order, so we only consider noise on $\epsilon_m$ here. The second-order effects matter for DFS encoding where the qubit is operated in the sweet spot, but our qubits are intentionally biased to the charge-sensitive RX regime where the gate time ($\lesssim 10$ ns for the parameters chosen in the main text) is significantly smaller than the timescale of second-order effects.

The average gate fidelity of the two-qubit gate, given the dephasing rate, is: \cite{Pedersen07, Harvey18}
\begin{eqnarray}
\text{IF}_D &=&1-\text{F}_D=\frac{4}{5}\left(\Gamma^D_2t_g\right)^2.
\end{eqnarray}
The definition and expressions for the average gate fidelity is given in Appendix~\ref{sec:a_F}.

For the CZ gate without dynamical decoupling, $t_g = \pi \sqrt{6}/\Delta \omega_c$. As $\Gamma^D_{2,0}(t)$ is only weekly dependent on time, we take the time average of $\Gamma^D_{2,0}(t)$, to yield:
\begin{eqnarray}
    \text{IF}_D \simeq \frac{96\pi^2}{5}A^D\left(\alpha^D\frac{\partial \omega_c}{\partial n^C_g}\right)^{-2}\mathbf{E}\left[\sqrt{\text{ln}\left(\frac{1}{\omega_l t}\right)}\right]^2,
\end{eqnarray}
where $\mathbf{E}\left[\sqrt{\text{ln}\left(1/\omega_lt)\right)}\right]$ is the time average of $\sqrt{\text{ln}\left(1/\omega_lt)\right)}$ over the gate time $[0, t_g]$. Here, we used the expression:
\begin{eqnarray}
    \frac{\partial \omega_q}{\partial \epsilon_m}
    \simeq \langle 1|\hat{n}^D_2|1\rangle^D-\langle 0|\hat{n}^D_2|0\rangle^D = -\frac{2\Delta n^C_g}{\alpha^D},
\end{eqnarray}
which comes from Eq.~\ref{eq:del_n_g_fh} and Eq.~\ref{eq:fh_qubit_sensitivity_2}.

For the CZ gate with dynamical decoupling, the qubit dephasing is dominant in the RX regime. This happens for twice the $\sqrt{\mathrm{CZ}}$ gate time: $2t_g=32/\Delta \omega_c$. In this case, the infidelity from the dephasing of qubits is:
\begin{eqnarray}
    \text{IF}_D \simeq \frac{16384}{5}A^D\left(\alpha^D\frac{\partial \omega_c}{\partial n^C_g}\right)^{-2}\text{ln}(2).
\end{eqnarray}

In both cases, given a fixed charge noise power on $\epsilon_m$, the infidelity from dephasing of qubits is only dependent on the lever arm and the sensitivity of the coupler, $\partial \omega_c/\partial n^C_g$. The smallest infidelity from the qubit dephasing is achieved when the charge sensitivity of the coupler is the highest. The change of quantum dot parameters does not affect the infidelity in first order, because the change in gate times and dephasing rates cancel each other for $1/f$ noise.

\subsection{\label{sec:4.2} OCS Transmon Coupler Dephasing}
We define the noise Hamiltonian $\hat{H}^C_n$ for the coupler similarly to the qubit noise Hamiltonian:
\begin{equation}
    \hat{H}^C_n = \zeta^C(t) \sum_{a,b=0,1}|abe\rangle\langle abe|.
\end{equation}

The $1/f$ noise on the coupler frequency is described by a double-sided noise spectral density $S[\omega]$:
\begin{eqnarray}
    S^C[\omega] = \begin{cases}
        A^C_\omega \frac{2\pi}{|\omega|} &\text{if $\omega_l < |\omega| <\omega_h$} \\
        0 &\text{otherwise}
    \end{cases},
\end{eqnarray}
where $A^C_\omega$ is the power spectral density at \SI{1}{Hz}, $\omega_l$ is the low-frequency cutoff, and $\omega_h$ is the high-frequency cutoff.

The noise on coupler frequency is induced by the noise on coupler gate charge:
\begin{eqnarray}
    A^C_\omega = \left(\frac{\partial \omega_c}{\partial n^C_g}\right)^2 \frac{1}{(2e)^2} A^C,
\end{eqnarray}
where $\partial \omega_c/\partial n^C_g$ is the charge sensitivity of the coupler and $A^C$ is the charge noise power spectral density on the coupler gate charge, at 1 Hz.

For a time-domain simulation, we first define $\hat{\mathcal{V}}(t)$, the coupler noise operator in the interaction picture with respect to the noise-free rotating frame Hamiltonian $\hat{H}_R(t)$:
\begin{eqnarray}
    \hat{\mathcal{V}}(t) = \hat{U}_0^\dagger (t, t_0)\left(\sum_{a,b=0,1}|abe\rangle\langle abe|\right) \hat{U}_0(t, t_0),
\end{eqnarray}
where $\hat{U}_0(t, t_0)$ is the evolution operator:
\begin{eqnarray}
    \hat{U}_0(t, t_0) = \mathcal{T} \exp\left(-i\int_{t_0}^t \hat{H}_R(t') dt'\right).
\end{eqnarray}
The time-ordering operator $\mathcal{T}$ is required because the noise-free Hamiltonian in general does not commute in time, i.e. $[\hat{H}_R(t_1), \hat{H}_R(t_2)]\neq 0$.

A truncated cumulant expansion provides a 2$^{\mathrm{nd}}$-order approximation of the system evolution~\cite{Kampen74, Groszkowski23}:
\begin{equation}
    \frac{\partial \hat{\rho}}{\partial t} = -\left[\hat{\mathcal{V}}(t), \left[\hat{\mathcal{V}}_{avg}(t), \hat{\rho}(t)\right] \right] ,\label{eq:time_evolution}
\end{equation}
where
\begin{eqnarray}
    \hat{\mathcal{V}}_{avg} = \int_0^t dt_1 S(t-t_1) \hat{\mathcal{V}}(t_1).
\end{eqnarray}
$S(t)$ is the autocorrelation function for $\zeta^C(t)$ in time domain, approximated as
\begin{eqnarray}
    S(t) \simeq -2A^C_\omega \mathrm{Ci}(\omega_l|t|)
\end{eqnarray}
and $\mathrm{Ci}(t) = -\int_t^\infty dt' \cos(t')/t'$, for a sufficiently large high-frequency cutoff $\omega_h$.

Numerical simulation of Eq.~\ref{eq:time_evolution} is used to calculate the final density matrix after evolution $\hat{\rho}(\hat{\rho}_i, t_g)$ for input states $\hat{\rho}_i \in \{\hat{\rho}_0\}$. These are compared to each ideal final state without decoherence to calculate the average gate fidelity, as shown in Appendix~\ref{sec:a_F}. Note that $\hat{\rho}(\hat{\rho}_i, t_g)$ takes into account coherent errors that arise from approximations involved in pulse-envelope-engineering.

As in the case of qubit dephasing, the circuit parameters $E_J$ and $E_C$ do not strongly affect the infidelity from coupler dephasing. This is seen through the numerical simulations, but it is also apparent analytically, as we next show. Suppose that a change in $E_J$ and $E_C$ changes the coupler sensitivity $\partial \omega_c/\partial n^C_g$ by a factor of $\gamma$:
\begin{eqnarray}
    \frac{\partial \omega_c}{\partial n^C_g} \rightarrow \gamma\frac{\partial \omega_c}{\partial n^C_g}.
\end{eqnarray}

This alters $\sigma^C$, $\Delta \omega_c$, and $t_g$:
\begin{eqnarray}
    A^C_\omega & \rightarrow & \gamma^2  A^C_\omega, \\
    \Delta \omega_c & \rightarrow &\gamma \Delta \omega_c, \\
    t_g & \rightarrow &t_g/\gamma.
\end{eqnarray}

Then $\hat{\mathcal{V}}(t)$ and $B(t)$ become:
\begin{eqnarray}
    \hat{\mathcal{V}}(t) & \rightarrow &\hat{\mathcal{V}}(\gamma t), \\
    \hat{\mathcal{V}}_{avg} & \rightarrow &\gamma \int_0^{\gamma t} dt_1 S\left(\frac{1}{\gamma}(\gamma t-t_1)\right) \hat{\mathcal{V}}(t_1).
\end{eqnarray}
For the $1/f$ noise we are considering, if the low-frequency cutoff $\omega_l$ is sufficiently smaller than $\Delta \omega_c$, $S(t/\gamma) \sim S(t)$. In this case, $\hat{\mathcal{V}}_{avg} \rightarrow \gamma \hat{\mathcal{V}}_{avg}(\gamma t)$. Thus,
\begin{eqnarray}
    \left. \frac{\partial \hat{\rho}}{\partial t} \right|_{t} \rightarrow \gamma \left. \frac{\partial \hat{\rho}}{\partial t} \right|_{\gamma t},
\end{eqnarray}
which ensures that the density matrix evolves through a similar trajectory during the gate, as long as the noise power is not strong. This can be intuitively explained as the change in noise power and the change in gate time cancelling out, similar as in the case of the spin qubit dephasing.

Other parameters do affect the infidelity from coupler dephasing. The higher the qubit sensitivity $\partial \omega_q/\partial \epsilon_m$ and the lever arm $\alpha^D$ are, the smaller $\text{IF}_C$ is. This is because $\Delta n^C_g$ increases and the noise power relative to $\Delta \omega_c$ decreases. Also, the increase in the noise power on the coupler $A^C$ leads to a trivial increase of $\text{IF}_C$.

\section{\label{sec:a_F} Definition of Average Gate Fidelity}
The average gate fidelity is found by the fidelity between the actual (noisy) and ideal (noiseless) final density matrices after the gate operation. To calculate the average gate fidelity, we first find the entanglement fidelity $F_e$, which is the averaged fidelity over the trace-orthonormal basis for all possible input density matrices, $\{\hat{\rho}_0\}$. Following other works, we tensor product single-qubits states $|0\rangle^D$, $|1\rangle^D$, $(|0\rangle^D + |1\rangle^D)/\sqrt{2}$, and $(|0\rangle^D + i|1\rangle^D)/\sqrt{2}$, with the coupler in ground state $|g\rangle$. This is expressed as~\cite{Nielsen02, Pedersen07, Harvey18}:
\begin{eqnarray}
    F_e = \frac{1}{16} \sum_{\hat{\rho}_i \in \{\hat{\rho}_0\}} \text{Tr}\left(\hat{\rho}(\hat{\rho}_i, t_g) \hat{\rho}_{\mathrm{ideal}}(\hat{\rho}_i)\right),
\end{eqnarray}
where
\begin{eqnarray}
    \{\hat{\rho}_0\} &= & \{\ket{\psi}\!\bra{\psi}_A \otimes \ket{\psi}\!\bra{\psi}_B \otimes |g\rangle\} \nonumber \\ && \quad \text{for} ~\ket{\psi}_A, \ket{\psi}_B \in \{|0\rangle, |1\rangle, 
 \nonumber \\ && \quad \quad  (|0\rangle + |1\rangle)/\sqrt{2}, (|0\rangle + i|1\rangle)/\sqrt{2}\}].
\end{eqnarray}
Here, $\hat{\rho}_{\mathrm{ideal}}=\hat{U}_{\mathrm{CZ}} \hat{\rho}_i \hat{U}_{\mathrm{CZ}}^\dagger$ is the ideal final density matrix after the gate operation, for the given input state $\hat{\rho}_i$, and $\hat{U}_{\mathrm{CZ}}$ is the unitary evolution operator for the CZ gate.

For a two-qubit gate, the relationship between the average gate fidelity and the entanglement fidelity 
is~\cite{Nielsen02, Pedersen07}:
\begin{eqnarray}
    F_g = \frac{1}{5}(4F_e+1).~\label{eq:gate_fidelity}
\end{eqnarray}
Eq.~\ref{eq:gate_fidelity} is a general form that could be applied to any gate operations, which are used in the time-domain simulations of Appendix~\ref{sec:4.2}. For the special case where the decoherence is dominated by a Gaussian decay with a certain decay rate $\Gamma$, the gate fidelity is:
\begin{eqnarray}
    F_g=1-\frac{4}{5}\left(\Gamma t_g\right)^2 \label{eq:dephasing_analytic},
\end{eqnarray}
which is the equation used in the analysis of Appendix~\ref{sec:4.1}. The derivation of Eq.~\ref{eq:dephasing_analytic} starts from the Kraus operators for single-qubit dephasing:
\begin{eqnarray}
G_0(t) = \begin{pmatrix}
1 & 0 \\
0 & \gamma(t)
\end{pmatrix}, \quad
G_1(t) = \begin{pmatrix}
0 & 0 \\
0 & \sqrt{1-\gamma(t)^2}
\end{pmatrix}, \label{eq:kraus_dephasing}
\end{eqnarray}
which could be easily verified:
\begin{equation}
\rho(t) = \sum_k G_k(t) \rho(0) G_k^\dagger(t) = \begin{pmatrix}
\rho_{00} & \gamma \rho_{10} \\
\gamma \rho_{01} & \rho_{11}
\end{pmatrix}.
\end{equation}
Here, $\gamma(t)$ is the decay factor for the off-diagonal elements of the density matrix. For a Gaussian decay profile (characteristic of the dephasing process under $1/f$ noise), $\gamma(t) = \exp\left(-t^2/T_2^2\right)$.

For two-qubit dephasing, assuming uncorrelated noise between the qubits, the Kraus operators for two-qubit dephasing are the tensor products of the single-qubit Kraus operators. With dephasing factors $\gamma^A$ and $\gamma^B$ corresponding to qubits $A$ and $B$, the Kraus operators are:

\begin{widetext}
\begin{eqnarray}
G_0 &=& G_0^A \otimes G_0^B = \begin{pmatrix}
1 & 0 & 0 & 0\\
0 & \gamma^B & 0 & 0 \\
0 & 0 & \gamma^A & 0 \\
0 & 0 & 0 & \gamma^A \gamma^B
\end{pmatrix}, \\
G_1 &=& G_0^A \otimes G_1^B = \begin{pmatrix}
1 & 0 & 0 & 0\\
0 & \sqrt{1-(\gamma^B)^2} & 0 & 0 \\
0 & 0 & \gamma^A & 0 \\
0 & 0 & 0 & \gamma^A \sqrt{1-(\gamma^B)^2}
\end{pmatrix}, \\
G_2 &=& G_1^A \otimes G_0^B = \begin{pmatrix}
1 & 0 & 0 & 0\\
0 & \gamma^B & 0 & 0 \\
0 & 0 & \sqrt{1-(\gamma^A)^2} & 0 \\
0 & 0 & 0 & \gamma^B \sqrt{1-(\gamma^A)^2}
\end{pmatrix}, \\
G_3 &=& G_1^A \otimes G_1^B = \begin{pmatrix}
1 & 0 & 0 & 0\\
0 & \sqrt{1-(\gamma^B)^2} & 0 & 0 \\
0 & 0 & \sqrt{1-(\gamma^A)^2} & 0 \\
0 & 0 & 0 & \sqrt{1-(\gamma^A)^2}\sqrt{1-(\gamma^B)^2}
\end{pmatrix}.
\end{eqnarray}
\end{widetext}

Finally, the relationship between Kraus operators and gate fidelity is given by~\cite{Pedersen07}:
\begin{equation}
F_g = \frac{1}{d(d+1)} \left[ \text{Tr}\left(\sum_k G_k^\dagger G_k\right) + \sum_k \left|\text{Tr}(G_k)\right|^2 \right], \label{eq:kraus_fidelity}
\end{equation}
where $d=4$ is the dimension of the two-qubit computational subspace.
Substituting $G_0$, $G_1$, $G_2$, and $G_3$ into Eq.~\ref{eq:kraus_fidelity} and assuming equal qubit dephasing rates $\Gamma = 1/T_2^A = 1/T_2^B$, Eq.~\ref{eq:dephasing_analytic} is recovered. Notably, Eq.~\ref{eq:dephasing_analytic} aligns with the equation for qubit dephasing infidelities in Ref.~\cite{Harvey18}, where the decay factor $\gamma$ is exponential due to the assumption of white noise.

\bibliography{mainbib}

\end{document}